\documentclass[10pt,a4paper,prd,tightenlines,nofootinbib,showpacs,showkeys,twocolomn, onecolomn]{revtex4}
\usepackage{bm}
\usepackage{graphics}
\usepackage{rotating}
\usepackage{epsfig}
\usepackage[usenames]{color}

\begin{document}

\title
{Weiss mean-field approximation for multicomponent stochastic spatially extended systems}
\author{\firstname{Svetlana E.} \surname{Kurushina}}
\affiliation{Physics Department, Samara State Aerospace University named after S.P. Korolyov, Moskovskoye Shosse 34, 443086,
Samara, Russian Federation}
\affiliation{Mathematics Department, Samara State Transport University, First Bezimyannii per., 18, 443066, Samara, Russian Federation}
\author{\firstname{Valerii V.} \surname{Maximov}}
\affiliation{Physics Department, Samara State Aerospace University named after S.P. Korolyov, Moskovskoye Shosse 34, 443086,
Samara, Russian Federation}
\affiliation{Mathematics Department, Samara State Transport University, First Bezimyannii per., 18, 443066, Samara, Russian Federation}
\author{\firstname{Yurii M.} \surname{Romanovskii}}
\affiliation{Physics Department, Lomonosov Moscow State University, GSP-1, Leninskie Gory, 119991,
Moscow, Russian Federation}

\begin{abstract}
We develop a mean-field approach for multicomponent stochastic spatially extended systems and use it to obtain a multivariate nonlinear self-consistent Fokker-Planck equation (NSCFPE) defining the probability density of the state of the system, which describes a well-known model of autocatalytic chemical reaction (brusselator) with spatially correlated multiplicative noise, and to study the evolution of probability density and statistical characteristics of the system in the process of spatial pattern formation. We propose the finite-difference method for numerical solving of a general class of multivariate nonlinear self-consistent time-dependent Fokker-Planck equations. We illustrate the accuracy and reliability of the method by applying it to an exactly solvable nonlinear Fokker-Planck equation (NFPE) for the Shimizu-Yamada model [Prog. Theor. Phys. \textbf{47}, 350 (1972)] and nonlinear Fokker-Planck equation [J. Stat. Phys. \textbf{19}, 1 (1978)] obtained for a nonlinear stochastic mean-field model introduced by Kometani and Shimizu [J. Stat. Phys. \textbf{13}, 473 (1975)]. Taking the problem indicated above as an example the accuracy of the method is compared with the accuracy of Hermite distributed approximating functional method [Phys. Rev. E \textbf{56}, 1197 (1997)]. Numerical study of the NFPE solutions for a stochastic brusselator shows that in the region of Turing bifurcation several types of solutions exist if noise intensity increases: unimodal solution, transient bimodality, and an interesting solution which involves multiple "repumping" of probability density through bimodality. Additionally we study the behavior of the order parameter of the system under consideration and show that the second type of solution arises in the supercritical region if noise intensity values are close to the values appropriate for the transition from bimodal stationary probability density for the order parameter to the unimodal one.
\end{abstract}



\maketitle

\section{Introduction}

Mean-field approximation (MFT) is an effective tool for the study of noise-driven dynamics of systems of different nature and noise-induced phenomena \cite{Lind1}. It is successfully applied for the study of noise-induced phase separation in conserved-order-parameter systems \cite{Iba2}, the noise-driven mechanism of pattern formation \cite{Buc3}, intrinsic noise-induced phase transitions \cite{Carr4}, non-equilibrium first-order phase transition induced by additive \cite{Zaik5} and multiplicative \cite{Mull6,Carr7} noise, noise-induced reentrant transition in nonlinear chains \cite{Land8}, pure noise-induced non-equilibrium second-order reentrant phase transition \cite{Bro9}, reentrant disorder-order-disorder and order-disorder-order phase transitions with the saddle-point structure of phase diagram \cite{Buce10}.

In quantum mechanics MFT implies the replacement of a multiparticle interaction Hamiltonian by a single-particle one. Weiss mean-field approximation for spatially extended systems implies that the interaction between a certain spatial point and its nearest neighbors occurs through the field, whose value corresponds to the statistically average field at this point. Herewith, a suitable way is used to carry out the discretization of the space of the initial spatially extended system and the Fokker-Planck equation (FPE) for the multivariate probability density function can be written for field values in the points of the obtained regular lattice. The obtained FPE is integrated over the values of the field at all points except the given one. This leads to FPE for the one-dimensional probability distribution density for field values at a given point. In the latter equation the conditional average values of the field at neighboring points are replaced by an average value of the field at a given point \cite{Iba2}.

Many real physical, chemical, biological, etc. systems are multi-component ones and they are modeled by means of partial differential equation systems. However, in \cite{Lind1,Iba2,Buc3,Carr4,Zaik5,Mull6,Carr7,Land8,Bro9,Buce10} only single-component spatially extended systems with additive, multiplicative or both noise types are considered. Therefore, one of the purposes of the present paper is to extend MFT for multicomponent stochastic reaction-diffusion systems which are a specific, but extremely important case of spatially extended systems.

Applying mean-field approximation to the study of noise induced phenomena arising in single-component problems leads to the necessity of numerical solution of a single-sight nonlinear self-consistent Fokker-Planck equation (NSCFPE). Various methods are used for the numerical integration of NSCFPE. In \cite{Zhan11,Zhang12} an elegant and effective method based on Hermite distributed approximating functionals is presented. High precision of the solution is achieved at small numbers of grid points. In \cite{Drozd13} the finite-difference method based on a K-point Stirling interpolation formula is proposed. In Ref. \cite{Chen14} a finite-difference scheme is used in the differential part and the trapezoid rule in the integral part of NFPE. Finite element \cite{Kum15} and finite-difference methods \cite{Kum15,Camp16}, discrete singular convolution algorithm \cite{Wei17}, direct quadrature based method of moments \cite{Ott18,Fox19}, pseudo-spectral method \cite{Kawam20}, path-integral \cite{Wehn21,Hak22} and eigenfunction expansion methods \cite{Kamp23,Tomit24} and others \cite{Mor25,Erm26} are also used to find numerical solutions of NFPEs.

Despite the variety of existing numerical methods of NFPE solution only a few of them, for example \cite{Zhang12,Kum15}, are successfully applied for the integration of multidimensional equations. Therefore numerical solution of multivariate NSCFPE is still a challenging problem, and the second purpose of the present paper is to propose the numerical method for this problem and to test its accuracy and reliability.

Finally, the third purpose is to apply the mean-field approach proposed and the method developed in Ref. \cite{Kur27} to the research of evolution of probability distribution density and statistical characteristics in the process of spatial pattern formation in the "brusselator" model \cite{Prig28}, which incorporates parameter fluctuations, and to compare the results of these approaches.

The rest of this paper is organized as follows. In Sec. II we introduce a generalized mean-field approach developed for multi-component stochastic reaction-diffusion systems and taking into account the spatial correlation of the external noise. In Sec. III the finite-difference method for numerically solving a general class of multivariate nonlinear self-consistent time-dependent Fokker-Planck equations is presented. The accuracy and reliability of the presented method are demonstrated by applying it to NSCFPE for the Shimizu-Yamada model \cite{Shim29,Shim30} and the Desai-Zwanzig model \cite{Desai31}. The results of comparing the accuracy of the proposed method with the accuracy of Hermite distributed approximating functional method \cite{Zhan11} are reviewed. Two-dimensional NSCFPE for spatially extended stochastic brusselator is derived in Sec. IV. Different types of this system probability density evolution arising with the noise intensity increase in the Turing bifurcation region are presented. The first and second order statistical characteristics of the system under consideration are studied. FPE for order parameters of the system under study is received. Its stationary solutions for the critical mode and its stationary statistical characteristics are explored. Finally, some conclusions are reported in Sec. V.

\section{MEAN-FIELD APPROACH FOR MULTICOMPONENT STOCHASTIC REACTION-DIFFUSION SYSTEMS}

The system of stochastic equations of the reaction-diffusion type is one of
the mathematical models describing the spatiotemporal dynamics of real
multi-component spatially extended systems under the influence of external
fluctuating environment and incorporating internal noises:

\begin{equation}
\label{eq1}
{\frac{{\partial x_{i}}} {{\partial t}}} = f_{i} (x_{1} ,...,x_{n} ) + g_{i}
(x_{1} ,...,x_{n} )\xi _{i} (\mathbf{r},t) + \eta _{i} (\mathbf{r},t)
+ D_{i} \nabla ^{2}x_{i} ,\,\,\,i = 1,...,n.
\end{equation}

\noindent
In Eq. (\ref{eq1}) $x_{i}$ are state functions of the system, $f_{i} (x_{1}
,...,x_{n} ),\,g_{i} (x_{1} ,...,x_{n} )$ are nonlinear functional
dependencies defining the interaction and evolution of components $x_{i}$ in
space and in time, $D_{i}$ are diffusion coefficients of components. The
additive random Gaussian fields $\eta _{i} (\mathbf{r},t)$ with zero means
and correlation functions $K{\left[ {\eta _{i} (\mathbf{r},t),\eta
_{{i}'} (\mathbf{{r}'},{t}')} \right]} = 2\zeta _{i} \delta (\mathbf{r} - \mathbf{{r}'})\delta (t - {t}')\delta _{i{i}'} $ model internal white noises, in the presence of which and in the absence of multiplicative
noise the system can exhibit equilibrium properties. The intensities of
internal noises are measured by parameters $\zeta _{i} $. Hereafter, we use
the notation $K[F_{1} ,F_{2} ]$ that is defined by the equality $K[F_{1}
,F_{2} ] = < F_{1} F_{2} > - < F_{1} > < F_{2} > $ for the correlation
function. The multiplicative random fields $\xi _{i} (\mathbf{r},t)$
model the external noises which disturb the system out of equilibrium. They
are also Gaussian \cite{Horst32} with zero means, but it is assumed that they are
homogeneous and spatio-isotropic and can have a nontrivial spatial
structure: $K{\left[ {\xi _{i} (\mathbf{r},t),\xi _{{i}'} (\mathbf{{r}'},{t}')} \right]} = 2\theta _{i} \Phi _{i} ({\left| {\mathbf{r} - \mathbf{{r}'}} \right|})\delta (t - {t}')\delta _{i{i}'} $, where $\Phi
_{i} ({\left| {\mathbf{r} - \mathbf{{r}'}} \right|})$ are spatial
correlation functions of external noises and $\theta _{i} $ are their
intensities. Further, to be definite, we shall use exponential spatial
correlation functions: $\Phi _{i} ({\left| {\mathbf{r} - \mathbf{{r}'}} \right|}) = \exp [ - k_{fi} ({\left| {\mathbf{r} - \mathbf{{r}'}} \right|})]$. Parameters $k_{fi} $ characterize the correlation
lengths $r_{fi} $ of noises: $r_{fi} = 1 / k_{fi} $.

In Ref. \cite{Iba2} the main aspects of mean-field approximation in application to
non-conserved systems with order parameter (model A in terms of literature
of critical phenomena) are outlined.

We carry out the discretization of continuous $d$-dimensional space of the system
(\ref{eq1}) and obtain a regular $d$-dimensional lattice with the mesh size $\Delta r$
and lattice points, the location of which will be characterized by vectors
${\rm {\bf r}}_{l} ,\,l = 1,...,p.$ Thus, regardless of the dimensionality
of the lattice each lattice point will correspond to only one index. We
assume that the interaction takes place only between the nearest neighbors,
which allows us to approximate the Laplace operator by a finite-difference
expression with a second-order difference. As a result of the
discretization the system (\ref{eq1}) is replaced by the system $n\times p$ of
ordinary differential equations:

\[
{\frac{{dx_{il}}} {{dt}}} = F_{il} (t),\,i = 1,...,n;\,l = 1,...,p,
\]

\begin{equation}
\label{eq2}
F_{il} (t) = f_{il} + g_{il} \xi _{il} (t) + \eta _{il} (t) + {\frac{{D_{i}
}}{{2d(\Delta r)^{2}}}}{\sum\limits_{{l}'} {\Lambda _{l{l}'}}}  x_{i{l}'}
.\,\,
\end{equation}

\noindent
In Eqs. (\ref{eq2}) the following notations are introduced: $f_{il} = f_{i} (x_{1l}
,...,x_{nl} ),\,\,g_{il} = g_{i} (x_{1l} ,...,x_{nl} ).$
${\sum\nolimits_{{l}'} {\Lambda _{l{l}'}}}  $ is the discrete analog of the
Laplace operator \cite{Iba2}: ${\sum\nolimits_{{l}'} {\Lambda _{l{l}'}}}   =
{\sum\nolimits_{{l}'} {(\delta _{nn(l),{l}'} - 2d\delta _{l,{l}'} )}} $,
where $nn(l)$ is a set of indexes of all sites which are the nearest
neighbors of the site with index $l$. The discrete noises $\eta _{il}
(t),\,\,\xi _{il} (t)$ have the correlation functions

\begin{equation}
\label{eq3}
 K{\left[ {\eta _{il} (t),\eta _{{i}'{l}'} ({t}')} \right]} = 2\zeta _{i}
{\frac{{\delta _{l{l}'}}} {{(\Delta r)^{d}}}}\delta (t - {t}')\delta
_{i{i}'}{\  \rm{and}\  }K{\left[ {\xi _{il} (t),\xi _{{i}'{l}'} ({t}')} \right]} =
2\theta _{i} \Phi _{i,\vert l - {l}'\vert}  \delta (t - {t}')\delta _{i{i}'}
\end{equation}

\noindent
Here, we have incorporated the fact that the continuum delta function
$\delta (\mathbf{r} - \mathbf{{r}'})$ has been replaced in the usual
way by a ratio that contains the Kronecker delta and the lattice spacing
$\delta _{l{l}'} / (\Delta r)^{d}$, and $\Phi _{i,\vert l - {l}'\vert}  $ is
convenient discretization of function $\Phi _{i} ({\left| {\mathbf{r} -
\mathbf{{r}'}} \right|})$. The values of $\Phi _{i,0} $ required further
can be computed numerically \cite{Garc33}.

The Fokker-Planck equation corresponding to equations (\ref{eq2}) in the
Stratonovich interpretation \cite{Strat34} for multivariate probability density
$\tilde {w}(x_{11} ,...,x_{1l} ,...,x_{1p} ,...,x_{n1} ,...,x_{nl}
,...,x_{np} ;t) = \tilde {w}(\{x_{1} ,...,x_{n} \};t)$ has the form:

\[
{\frac{{\partial \tilde {w}(\{x_{1} ,...,x_{n} \};t)}}{{\partial t}}} = -
{\sum\limits_{i = 1}^{n} {{\sum\limits_{{l}' = 1}^{p} {{\frac{{\partial
}}{{\partial x_{i{l}'}}} }}} {\left\{ {\left( {{\left\langle {F_{i{l}'} (t)}
\right\rangle}  + {\sum\limits_{j = 1}^{n} {{\sum\limits_{m = 1}^{p}
{{\int\limits_{ - \infty} ^{0} {K{\left[ {{\frac{{\partial F_{i{l}'}
(t)}}{{\partial x_{jm}}} },F_{jm} (\tau )} \right]}d\tau}} } }} }}
\right)\tilde {w}} \right\}}}}  +
\]
\begin{equation}
\label{eq4}
 + {\sum\limits_{i,j = 1}^{n} {{\sum\limits_{m,{l}' = 1}^{p}
{{\frac{{\partial ^{2}}}{{\partial x_{i{l}'} x_{jm}}} }{\left\{ {\left(
{{\int\limits_{ - \infty} ^{0} {K{\left[ {F_{i{l}'} (t),F_{jm} (\tau )}
\right]}d\tau}} }  \right)\tilde {w}} \right\}}}}} } .
\end{equation}

for all lattice points.

Considering Eqs. (\ref{eq2}) the correlators included in Eq. (\ref{eq4}) are easily
computed:

\[
K{\left[ {{\frac{{\partial F_{i{l}'} (t)}}{{\partial x_{jm}}} },F_{jm} (\tau
)} \right]} = {\frac{{\partial g_{i{l}'}}} {{\partial x_{jm}}} }g_{jm} K[\xi
_{i{l}'} (t),\xi _{jm} (\tau )]\delta _{ij} \delta _{m{l}'} ,m = l,nn(l),
\]

\begin{equation}
\label{eq5}
K{\left[ {F_{i{l}'} (t),F_{jm} (\tau )} \right]} = \left( {g_{i{l}'} g_{jm}
K[\xi _{i{l}'} (t),\xi _{jm} (\tau )] + K[\eta _{i{l}'} (t),\eta _{jm} (\tau
)]} \right)\delta _{ij} \delta _{m{l}'} ,m = l,nn(l).
\end{equation}

After the substitution of correlators (\ref{eq5}) and (\ref{eq3}) into Eq. (\ref{eq4}) and some
simple transformations the equation for multivariate probability
density $\tilde {w}(\{x_{1} ,...,x_{n} \};t)$ will appear as:

\[
{\frac{{\partial \tilde {w}(\{x_{1} ,...,x_{n} \};t)}}{{\partial t}}} =
\]
\begin{equation}
\label{eq6}
- {\sum\limits_{i = 1}^{n} {{\sum\limits_{{l}' = 1}^{p} {{\frac{{\partial
}}{{\partial x_{i{l}'}}} }}}} } {\left[ {f_{i{l}'} + {\frac{{D_{i}
}}{{2d(\Delta r)^{2}}}}\left( {{\sum\limits_{m = nn({l}')} {x_{im}}}   -
2dx_{i{l}'}}  \right) -  \\
{\sum\limits_{m = {l}',nn({l}')} {\left( {\zeta _{i}
{\frac{{\partial}} {{\partial x_{im}}} } - \theta _{i} g_{i{l}'} \Phi
_{i,\vert {l}' - m\vert}  {\frac{{\partial}} {{\partial x_{im}}} }g_{im}}
\right)}}}  \right]}\tilde {w}.
\end{equation}

We choose one site with index $l$. In order to obtain multivariate probability
density $w(x_{1l} ,...,x_{il} ,...,x_{nl} ;t) = w(\{x\};t)$ for a single
site it is necessary to integrate $\tilde {w}(x_{11} ,...,x_{1l}
,...,x_{1p},...,x_{n1},...,x_{nl},...,x_{np};t)$ over all the variables except $x_{1l} ,...,x_{il}
,...,x_{nl} $:

\[
w(x_{1l} ,...,x_{il} ,...,x_{nl} ;t) = \int {\tilde {w}(x_{11} ,...,x_{1l}
,...,x_{1p} ,...,x_{n1} ,...,x_{nl} ,...,x_{np} ;t){\left[ {{\prod\limits_{k
\ne l} {dx_{1k}}}  ...dx_{ik} ...dx_{nk}}  \right]}} .
\]

We use the property of probability density to vanish at the infinity:
$\tilde {w}(\{x_{1} ,...,x_{n} \};t) \to 0$ if $x_{il} \to \pm \infty ,\,i =
1,...,n;l = 1,...,p.$ Then

\begin{equation}
\label{eq7}
\int {{\frac{{\partial}} {{\partial x_{im}}} }\left( {g_{im} \tilde {w}}
\right){\left[ {{\prod\limits_{k \ne l} {dx_{1k}}}  ...dx_{ik} ...dx_{nk}}
\right]}} = {\left\{ {{\begin{array}{*{20}c}
 {0,m \ne l,} \hfill \\
 {{\frac{{\partial}} {{\partial x_{il}}} }{\left[ {g_{il} w(\{x\};t)}
\right]},m = l.} \hfill \\
\end{array}}}  \right.}
\end{equation}

According to the definition of conditional probability we can write:

\[
\int {x_{im} \tilde {w}{\left[ {{\prod\limits_{k \ne l} {dx_{1k}}
}...dx_{ik} ...dx_{nk}}  \right]}} = \int {x_{im} w(x_{1l} ,...,x_{im}
,x_{il} ,...,x_{nl} ;t)dx_{im}}  =
\]

\begin{equation}
\label{eq8}
 = {\left[ {\int {x_{im} w(x_{im} \vert x_{1l} ,...,x_{il} ,...,x_{nl}
;t)dx_{im}}}   \right]}w(x_{1l} ,...,x_{il} ,...,x_{nl} ;t) =
w(\{x\};t)E(x_{im} \vert x_{1l} ,...,x_{il} ,...,x_{nl} ;t).
\end{equation}

Here $E(x_{im} \vert x_{1l} ,...,x_{il} ,...,x_{nl} ;t)$ are
nearest-neighbor conditional averages.

Finally, taking into account Eqs.(\ref{eq7},\ref{eq8}), we get:
\[
{\frac{{\partial w(\{x\};t)}}{{\partial t}}} =
\]
\begin{equation}
\label{eq9}
 - {\sum\limits_{i = 1}^{n} {{\frac{{\partial}} {{\partial x_{il}}} }}
}{\left[ {f_{il} + {\frac{{D_{i}}} {{2d(\Delta r)^{2}}}}\left(
{{\sum\limits_{m = nn(l)} {E(x_{im} \vert x_{1l} ,...,x_{il} ,...,x_{nl}
;t)}}  - 2dx_{il}}  \right) - \zeta _{i} {\frac{{\partial}} {{\partial
x_{il}}} } - \theta _{i} \Phi _{i,0} g_{il} {\frac{{\partial}} {{\partial
x_{il}}} }g_{il}}  \right]}w
\end{equation}

\noindent
for single-point multivariate probability density.

Taking into account that $x_{il} $ are linked by the equations (\ref{eq2}) let us
assume that the mean field approximation is to imply that the conditional
average $E(x_{im} \vert x_{1l} ,...,x_{il} ,...,x_{nl} ;t)$ in eq. (\ref{eq9}) can
be replaced by the conditional average $E({\left. {x_{il}}  \right|}x_{1l}
,...,x_{i - 1l} ,x_{i + 1l} ,...,x_{nl} ;t)$:

\begin{equation}
\label{eq10}
E(x_{im} \vert x_{1l} ,...,x_{il} ,...,x_{nl} ;t) = E({\left. {x_{il}}
\right|}x_{1l} ,...,x_{i - 1l} ,x_{i + 1l} ,...,x_{nl} ;t),
\end{equation}

\begin{equation}
\label{eq11}
\begin{array}{l}
 E({\left. {x_{il}}  \right|}x_{1l} ,...,x_{i - 1l} ,x_{i + 1l} ,...,x_{nl}
;t) = {\int\limits_{ - \infty} ^{ + \infty}  {x_{il}}}  w({\left. {x_{il}}
\right|}x_{1l} ,...,x_{i - 1l} ,x_{i + 1l} ,...,x_{nl} ;t)dx_{il} , \\
 w({\left. {x_{il}}  \right|}x_{1l} ,...,x_{i - 1l} ,x_{i + 1l} ,...,x_{nl}
;t) = {\frac{{w(\{x\};t)}}{{{\int\limits_{ - \infty} ^{ + \infty}  {w(x_{1l}
,...,x_{il} ,...,x_{nl} ;t)dx_{il}}} } }}. \\
 \end{array}
\end{equation}

In this approximation the exact FPE (\ref{eq9}) is transformed into an approximate
\[
{\frac{{\partial w(\{x\};t)}}{{\partial t}}} =
\]
\begin{equation}
\label{eq12}
- {\sum\limits_{i = 1}^{n} {{\frac{{\partial}} {{\partial x_{il}}} }}
}{\left[ {f_{il} + {\frac{{D_{i}}} {{(\Delta r)^{2}}}}\left( {E({\left.
{x_{il}}  \right|}x_{1l} ,...,x_{i - 1l} ,x_{i + 1l} ,...,x_{nl} ;t) -
x_{il}}  \right) - \zeta _{i} {\frac{{\partial}} {{\partial x_{il}}} } -
\theta _{i} \Phi _{i,0} g_{il} {\frac{{\partial}} {{\partial x_{il}
}}}g_{il}}  \right]}w.
\end{equation}

Hereafter the index $l$ is omitted for the simplicity of writing.

Equations (\ref{eq10})-(\ref{eq12}) form a self-consistent system for which it is impossible
to write a stationary solution even implicitly as opposed to the
one-dimensional case. The numerical solution of (\ref{eq10})-(\ref{eq12}) is a complicated
problem. The next section will be devoted to the development of a numerical
method for the solution of this problem and testing its accuracy and
reliability.

\section{NUMERICAL METHOD FOR THE MULTIVARIATE NONLINEAR SELF-CONSISTENT FOKKER-PLANCK EQUATION}

\subsection{Finite-difference method}

A multivariate nonlinear self-consistent Fokker-Planck equation (\ref{eq12}) can be
presented as:

\begin{equation}
\label{eq13}
{\frac{{\partial w}}{{\partial t}}} = {\sum\limits_{\alpha = 1}^{n}
{{\frac{{\partial}} {{\partial x_{\alpha}} } }\left( {k_{\alpha}
(x,t){\frac{{\partial w}}{{\partial x_{\alpha}} } } - r_{\alpha}  (x,t)w}
\right)}}  = {\sum\limits_{\alpha = 1}^{n} {L_{\alpha}  w}} ,	x = (x_{1}
,...,x_{n} ),	\alpha = 1,...,n,
\end{equation}

\noindent
where $k_{\alpha} (x,t) = \zeta _{\alpha}  + g_{\alpha} ^{2} \theta
_{\alpha}  \Phi _{\alpha ,0} ,	k_{\alpha}  (x,t) > 0,$

\noindent
$
r_{\alpha}  (x,t) = - f_{\alpha}  - {\frac{{D_{\alpha}} } {{(\Delta
r)^{2}}}}[E({\left. {x_{\alpha}}   \right|}x_{1} ,...,x_{\alpha - 1}
,x_{\alpha + 1} ,...,x_{n} ;t) - x_{\alpha}  ] + \theta _{\alpha}  \Phi
_{\alpha ,0} g_{\alpha}  {\frac{{\partial g_{\alpha}} } {{\partial x_{\alpha
}}} }.
$

Here $x = (x_{1} ,...,x_{n} )$ belong to the region $G$. The functions
$r_{\alpha}(x,t)$ implicitly depend on $w$ (see Eq.(\ref{eq11})).

Let us choose natural boundary conditions (BC) for the probability density:

\begin{equation}
\label{eq14}
 w(x,t) \to 0\ \ {\rm{if}}\ \ x_{\alpha}  \to \pm \infty
\end{equation}
\noindent
and the initial condition (IC)
\begin{equation}
\label{eq15}
w(x,0) = w_{0}(x).
\end{equation}

Let us transform the operators $L_{\alpha} $ to the form:

\begin{equation}
\label{eq16}
L_{\alpha}  = {\frac{{\partial}} {{\partial x_{\alpha}} } }\left(
{{\frac{{k_{\alpha}} } {{q_{\alpha}} } }{\frac{{\partial}} {{\partial
x_{\alpha}} } }(q_{\alpha}  w)} \right), \ \ \ q_{\alpha}  = \exp
\int {{\frac{{r_{\alpha}} } {{k_{\alpha}} } }} dx_{\alpha}  .
\end{equation}

The functions $q_{\alpha}  $ from (\ref{eq16}) obtained by integrating over
$x_{\alpha}  $ include the conditional average $E({\left. {x_{\alpha}}
\right|}x_{1} ,...,x_{\alpha - 1} ,x_{\alpha + 1} ,...,x_{n} ;t)$ that
represents the function of the variables $x_{1} ,...,x_{\alpha - 1}
,x_{\alpha + 1} ,...,x_{n} $ except $x_{\alpha}  $. Finding $q_{\alpha}  $,
therefore, does not present a problem. If $\int {(r_{\alpha}  /}  k_{\alpha
} )dx_{\alpha}  $ cannot be integrated precisely one can use approximative
methods, for example, the trapezoid rule.

For the problem (\ref{eq13})-(\ref{eq15}) we choose a rectangular spatial mesh $\omega _{h}
= (\{x_{i} \} = \{i_{1} h_{1} ,...,i_{\alpha}  h_{\alpha}  ,...,i_{n} h_{n}
\} \in G)$ where $i_{1} ,...,i_{n} \,\,\,(i_{\alpha}  = 0,1,...,N_{\alpha}
)$ and $h_{1} ,...,h_{n} $ are the indices of the mesh points and the steps
respectively, and $\omega _{\tau}  $ a time mesh with a step $\tau $ over
the interval $0 \le t \le T.$ For the mesh functions given on $\omega _{h}
\times \omega _{\tau}  $ we shall use the following notations:

\noindent
$
y = y^{j + \alpha / n} = y(x_{i} ,t_{j + \alpha / n} ),\,
$

\noindent
 $\,y_{\bar {x}_{\alpha}}   = [y(x_{1,i_{1}}  ,...,x_{\alpha ,i_{\alpha}}
,...,x_{n,i_{n}}  ,t) - y(x_{1,i_{1}}  ,...,x_{\alpha ,i_{\alpha}  - 1}
,...,x_{n,i_{n}}  ,t)] / h_{\alpha}  $ is the left-side difference derivative
at the point $x_{1,i_{1}}  ,...,x_{\alpha ,i_{\alpha}}   ,...,x_{n,i_{n}}
$,

\noindent
 $\,y_{x_{\alpha}}   = [y(x_{1,i_{1}}  ,...,x_{\alpha ,i_{\alpha}  + 1}
,...,x_{n,i_{n}}  ,t) - y(x_{1,i_{1}}  ,...,x_{\alpha ,i_{\alpha}}
,...,x_{n,i_{n}}  ,t)] / h_{\alpha}  $ is the right-side difference derivative
at the point $x_{1,i_{1}}  ,...,x_{\alpha ,i_{\alpha}}   ,...,x_{n,i_{n}}
$.

Applying the finite-volume method \cite{Samar35} we associate $L_{\alpha} $ to
difference analogs \cite{Karet36}:

\begin{equation}
\label{eq17}
\Lambda _{\alpha}  y = (a_{\alpha}  (q_{\alpha}  y)_{\bar {x}_{\alpha}}
)_{x_{\alpha}}   ,\,\,
\end{equation}

\noindent
here $\,a_{\alpha ,i} = {\left[ {{\int\limits_{x_{\alpha ,i_{\alpha}  - 1}
}^{x_{\alpha ,i_{\alpha}} }   {{\frac{{q_{\alpha}} } {{k_{\alpha}
}}}dx_{\alpha}} } }  \right]}^{ - 1}\,\,.$

A locally one-dimensional scheme for the problem (\ref{eq13})-(\ref{eq15}) will take the
form:

\begin{equation}
\label{eq18}
{\frac{{y^{j + \alpha / n} - y^{j + (\alpha - 1) / n}}}{{\tau}} } - \Lambda
_{\alpha}  y = 0,\,\,\,y^{0} = w_{0} .\,\,\,
\end{equation}

It is shown in Refs. \cite{Karet36,Samar37} that scheme (\ref{eq18}) is unconditionally stable in
the Banach space with the norm

\noindent
${\left\| {y} \right\|} = {\sum\limits_{i_{1}
}^{N_{1} - 1} {}} ...{\sum\limits_{i_{n}} ^{N_{n} - 1} {{\left|
{y(x_{1,i_{1}}  ,...,x_{\alpha ,i_{\alpha}}   ,...,x_{n,i_{n}}  )}
\right|}h_{1} ...h_{n}}}  $ and has the accuracy $O(\tau +
{\sum\nolimits_{\alpha}  {h_{\alpha} ^{2}}}  )$.

Depending on the sign of the function $r_{\alpha} (x,t)$ we can use the
appropriate variant of a tridiagonal matrix algorithm or any other method of
solving systems of linear algebraic equations. The integrals (\ref{eq11}) are easily
calculated using the Simpson's rule from the preceding layer.

\subsection{Accuracy, reliability, and limitations}

To demonstrate the accuracy and reliability of the results obtained with the
help of the finite-difference scheme (\ref{eq18}) and to find out its limitations we
apply (\ref{eq18}) to the exactly solvable NFPE for the Shimizu--Yamada model
\cite{Shim29,Shim30} and the NFPE \cite{Desai31} obtained for the nonlinear stochastic mean-field
model introduced by Kometani and Shimizu \cite{Komet38}. The accuracy of the method is
compared with the accuracy of Hermite distributed approximating functional
method \cite{Zhan11} taking the problem indicated above as an example.

NSCFPE for Shimizu--Yamada model has the form:

\begin{equation}
\label{eq19}
{\frac{{\partial f(x,t)}}{{\partial t}}} = {\frac{{\partial \{[\omega x +
\theta Ex(t)]f(x,t)\}}}{{\partial x}}} + D{\frac{{\partial
^{2}f(x,t)}}{{\partial x^{2}}}},
\end{equation}

\noindent
here $Ex(t) = {\int\limits_{ - \infty} ^{ + \infty}  {xf(x,t)}} dx$ is
mathematical expectation, $\omega ,\,\,\,\theta ,$ and $D$ are constants.
With the initial spatial distribution in the form of the Dirac delta
function $f(x,0) = \delta (x - x_{0} )$ the exact solution of the problem
(\ref{eq19}) takes the form (Fig.1)

\begin{equation}
\label{eq20}
f(x,t) = {\frac{{1}}{{\sqrt {2\pi \sigma (t)}}} }\exp {\left[ { - {\frac{{[x
- Ex(t)]^{2}}}{{2\sigma (t)}}}} \right]},
\end{equation}

\noindent
where $Ex(t) = x_{0} e^{ - (\omega + \theta )t},\,\,\,\sigma (t) =
{\frac{{D}}{{\omega}} }(1 - e^{ - 2\omega t}).$

\begin{figure}[!h]
\centering
\includegraphics[width=3.25in,height=3.18in]{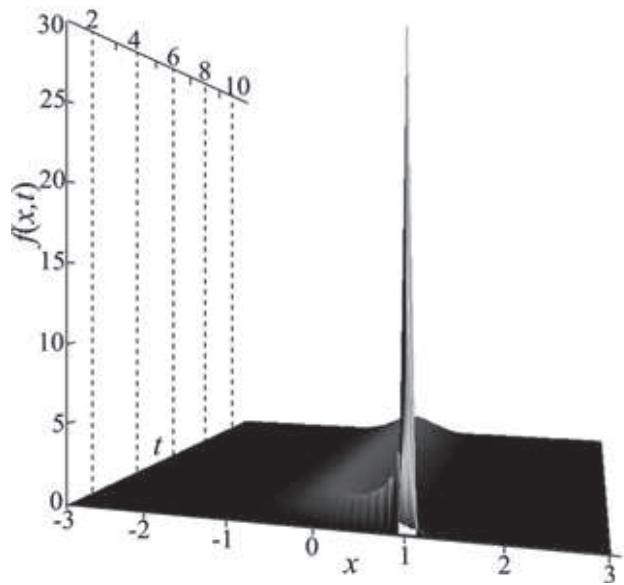}\\
\caption{ Analytical solution (\ref{eq20}) for the problem (\ref{eq19}). Hereafter the
parameters $\omega = 1$, $\theta = 1$, $D = 0.1$, $x_{0} = 1$ are chosen for
the problem (\ref{eq19}).}
\end{figure}

The accuracy and reliability of the scheme (\ref{eq18}) was determined by the
relative error
\begin{equation}
\label{eq21}
\varepsilon (t) = {\frac{{{\left. {\left( {Ex^{2}(t) -
[Ex(t)]^{2}} \right)} \right|}_{Num}}} {{\sigma (t)}}} - 1.
\end{equation}

\noindent
The expression ${\left. {\left( {Ex^{2}(t) - [Ex(t)]^{2}} \right)}
\right|}_{Num} $ is the second moment obtained in solving (\ref{eq19}) by various
numerical methods. Additional control was accomplished by checking the
fulfillment of the condition of probability density normalization per unit
${\rm l}(t) = {\int\limits_{ - \infty} ^{ + \infty}  {f(x,t)}} dx$ at each
time step. The condition was also used as the criterion of the correct
choice of the integration region size governing the observation of boundary
conditions (\ref{eq14}).

Figs. 2, 3 compare the plots of the decimal logarithm of the relative error
$\lg \vert \varepsilon (t)\vert $ and probability normalization ${\rm
l}(t)$. Fig.4 and Fig.5 demonstrate numerical solutions of eq. (\ref{eq19}) obtained
on the basis of the scheme (\ref{eq18}) and the method based on DAF. Fig.6 shows the
dependencies $f(x)$ for $t$ = 0.01 for the analytical solution and the solutions
obtained by methods (\ref{eq18}) and DAF. The appropriate dependencies are obtained
for the recommended (giving the least error) in \cite{Zhan11} parameters of the
DAF-based method in the proper order of time approximation ($O(\tau ))$, the
same as in scheme (\ref{eq18}). The solution by the scheme (\ref{eq18}) is obtained for
$h_{1} = \tau = 0.001$.

\begin{figure}[!h]
\centering
\includegraphics[width=3.25in,height=1.77in]{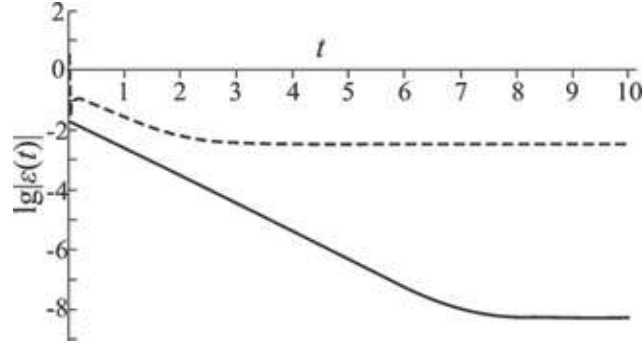}\\
\caption{Dependencies of the decimal algorithm of the relative error $\lg \vert \varepsilon (t)\vert $ (\ref{eq21}) on time. The dashed line is the DAF-based method \cite{Zhan11}, the solid line is the finite-difference method (\ref{eq18}).}
\end{figure}

\begin{figure}[!h]
\centering
\includegraphics[width=3.25in,height=1.83in]{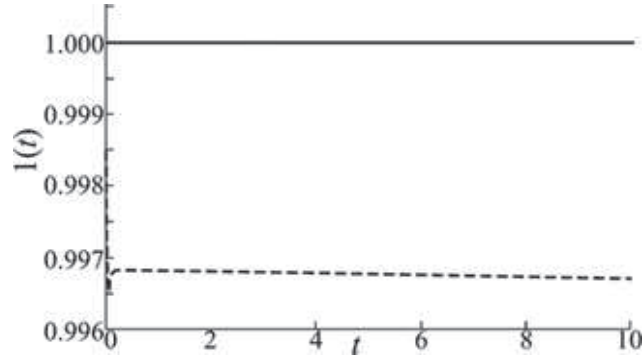}\\
\caption{Dependencies of probability density normalization ${\rm l}(t)$ on
time. The dashed line is the DAF-based method, the solid line is the
finite-difference method (\ref{eq18}).}
\end{figure}

\begin{figure}[!h]
\centering
\includegraphics[width=3.29in,height=3.12in]{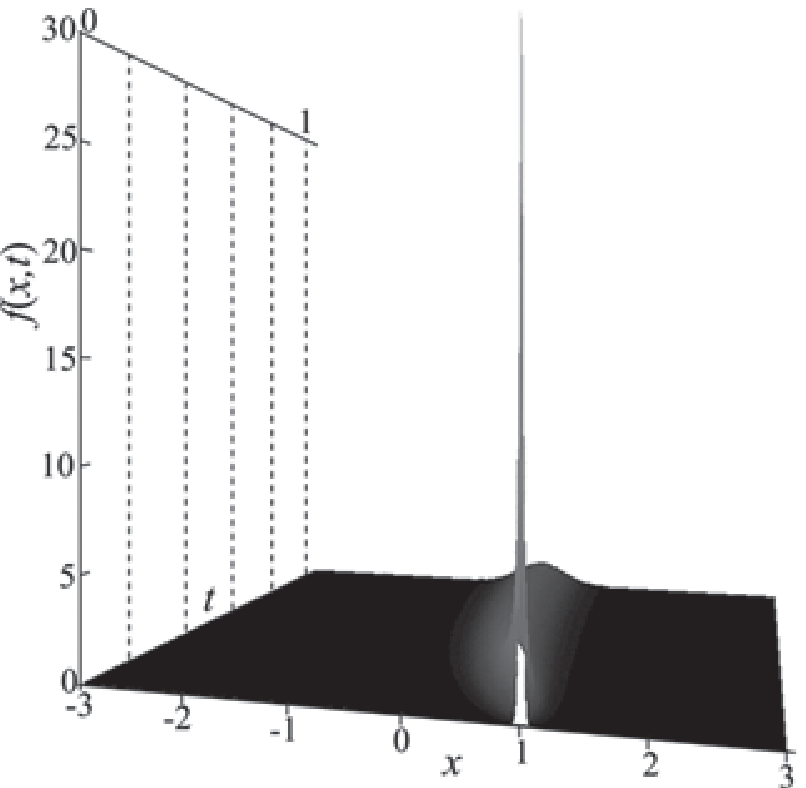}\\
\caption{Surface $f(x,t)$ obtained as a result of the numerical solution (\ref{eq19}) by
the method (\ref{eq18}).}
\end{figure}

\begin{figure}[!h]
\centering
\includegraphics[width=3.32in,height=2.20in]{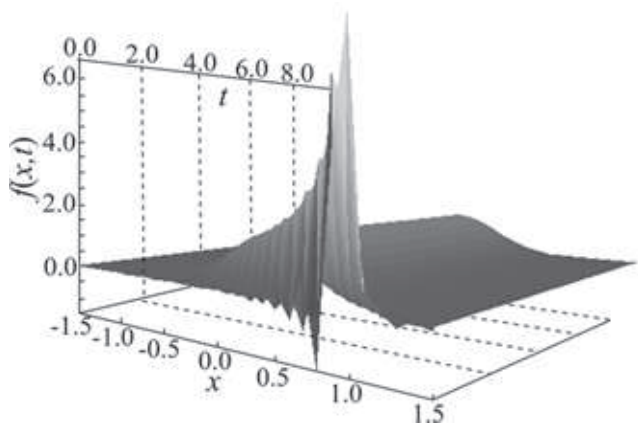}\\
\caption{Surface $f(x,t)$ obtained as a result of the numerical solution (\ref{eq19}) by
the DAF-based method.}
\end{figure}

\begin{figure}[!h]
\centering
\includegraphics[width=6.10in,height=2.68in]{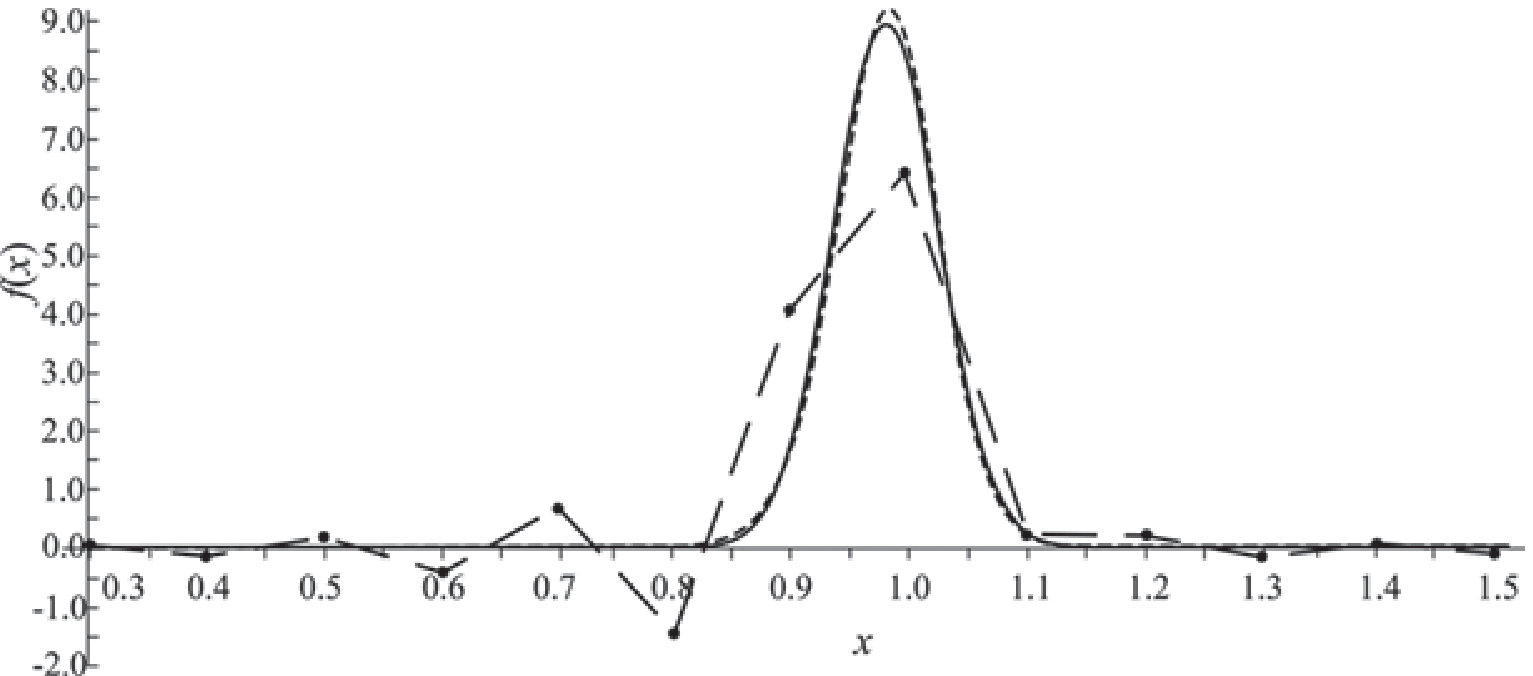}\\
\caption{The dependence of probability density $f(x)$ at $t = 0.01$. The solid line
is the analytical solution (\ref{eq20}), the dashed line is the solution obtained by
method (\ref{eq18}), and the line with a long dash is the DAF--based method \cite{Zhan11}.}
\end{figure}

It can be seen from the plots presented in Fig. 2 that in the first order of
approximation over time the accuracy of the solution obtained by using
scheme (\ref{eq18}) is higher, especially when steady-state values are reached and
time moments are close to the initial one. Fig. 3 shows that the dependence
${\rm l}(t)$ obtained by the DAF-based method is slowly decreasing, which
means the violation of the normalization condition. On the contrary, scheme
(\ref{eq18}) conserves asymptotically (at large times) the condition of
normalization of probability density per unit.

The analysis of Figs. 1, 4, 5, and 6 shows that scheme (\ref{eq18}) provides the
positive definiteness of probability density values at time moments close to
the initial one, where the solution is close to discontinuity.

The second problem chosen for the examination of scheme (\ref{eq18})

\begin{equation}
\label{eq22}
{\frac{{\partial f(x,t)}}{{\partial t}}} = {\frac{{\partial \{[x^{3} +
(\theta - 1)x - \theta E(x(t))]f(x,t)\}}}{{\partial x}}} + D{\frac{{\partial
^{2}f(x,t)}}{{\partial x^{2}}}}
\end{equation}
\noindent
has a larger computational complexity than problem (\ref{eq19}), since long-lived
bimodality is observed here at certain parameters, and consequently
calculations are to be performed for large times. It imposes an additional
requirement - asymptotic stability - on the numerical method.

The problem (\ref{eq22}) was solved numerically in Ref. \cite{Zhan11} by Hermite distributed
approximating functional method. Below the distributions $f(x)$ at different
time moments, the dependencies of normalization $1(t)$ on time, mean
$E(x(t))$, variance $Dx(t) = Ex^{2}(t) - [Ex(t)]^{2}$, and the decimal
logarithm of relative error $\lg \vert \varepsilon _{num} (t)\vert =
{\frac{{Dx(t)\vert _{Eq.18}}} {{Dx(t)\vert _{DAF}}} } - 1$ are presented in
Figs. 7-11 for comparison. In the last expression $Dx(t)\vert _{Eq.18} $ is
the variance calculated on the basis of scheme (\ref{eq18}), $Dx(t)\vert _{DAF} $ is
the variance calculated on the basis of the DAF-based method. Here the DAF-based
method solution is chosen as a benchmark for comparison.

\begin{figure}
\centering
\includegraphics[width=3.15in,height=2.75in]{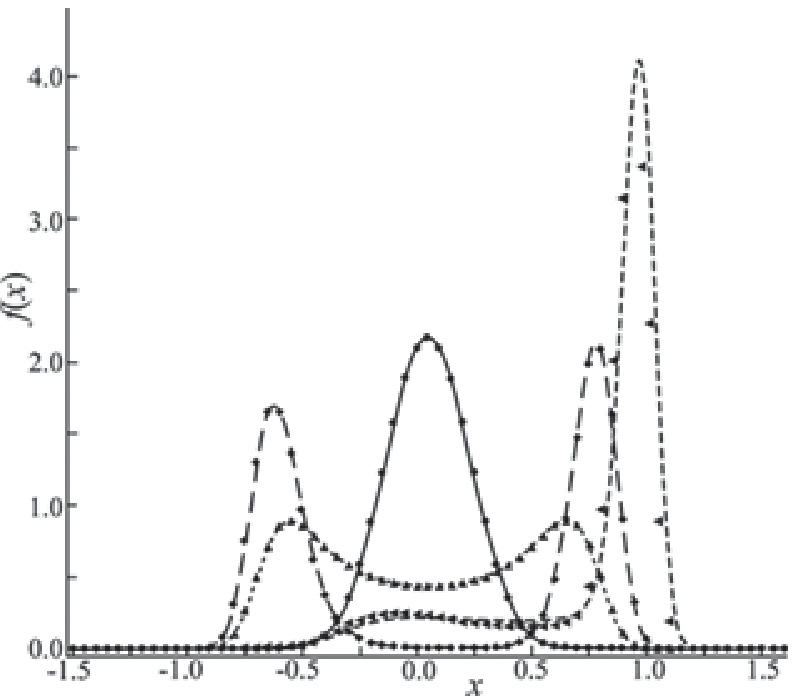}\\
\caption{Distributions $f(x)$ at time moments $t$ = 1, 4, 72, 105.5 obtained on the
basis of scheme (\ref{eq18}) (lines) and on the basis of the DAF-method (separate
symbols). $t$=1: solid line, circles; $t$=4: dotted line, triangles; $t$=72: line
with long dash, squares; $t$=105.5: dashed line, rotated triangles.
Hereinafter, the parameters $\theta = 0.5$, $D = 0.01$, $x_{0} = 10^{ - 4}$
are chosen for the problem (\ref{eq22}).}
\end{figure}

\begin{figure}
\centering
\includegraphics[width=3.04in,height=1.49in]{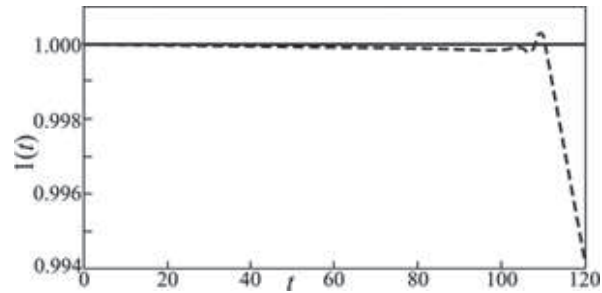}\\
\caption{Dependencies of normalization ${\rm l}(t)$ on time for the problem (\ref{eq22}). The dashed line is the DAF-based method \cite{Zhan11}, the solid line is the finite-difference method (\ref{eq18}).}
\end{figure}

\begin{figure}
\centering
\includegraphics[width=2.96in,height=1.62in]{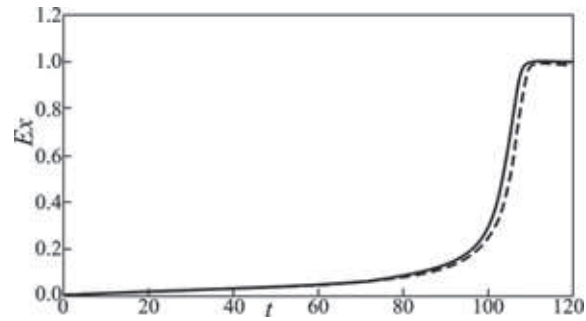}\\
\caption{Expectation $Ex(t)$ vs time. The dashed line is the DAF-based method
\cite{Zhan11}, the solid line is the finite-difference method (\ref{eq18}).}
\end{figure}

\begin{figure}
\centering
\includegraphics[width=2.89in,height=1.56in]{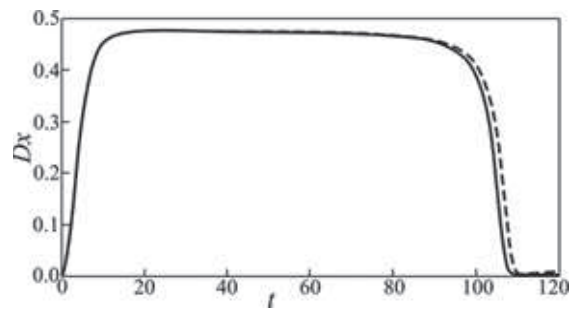}\\
\caption{Variance $Dx(t)$ vs time. The dashed line is the DAF-based method
\cite{Zhan11}, the solid line is the finite-difference method (\ref{eq18}).}
\end{figure}

\begin{figure}
\centering
\includegraphics[width=2.94in,height=1.71in]{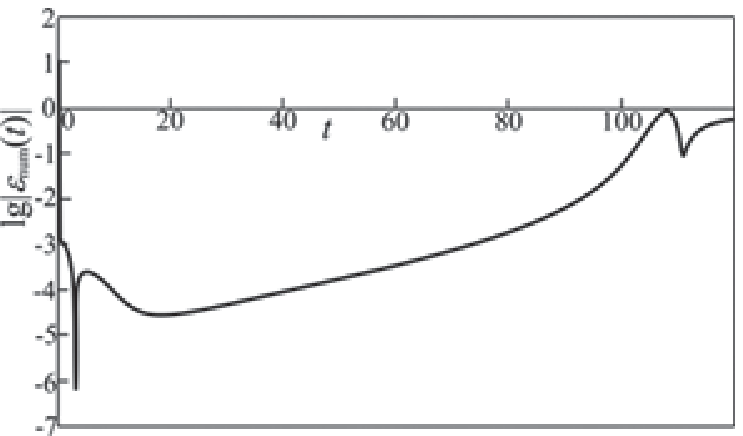}\\
\caption{Decimal logarithm of relative error $\lg \vert \varepsilon _{num}(t)\vert $ vs time for the problem (\ref{eq22}).}
\end{figure}

It can be seen from Fig. 7 that the difference between numerical solutions
begins to appear at times of the order of 105.5. Fig.8 gives the explanation
for this. It follows from it that significant violation of the normalization
condition arises at the same time. Evidently, insignificant differences in
dependencies of means $Ex(t)$ and variances $Dx(t)$ (see Figs. 9,10) and the
increase of relative error (see Fig. 11) are associated with it.

Thus, the DAF-based method limitations connected with the violation of
probability density normalization condition manifest themselves stronger in
the problems with large computational complexity. That is why its stability
and the reliability of solution fail at large times. Also this
method violates locally the positive definiteness of solutions required for
the fulfillment of the standard properties of the probability density. It
should be mentioned that for problem (\ref{eq13})-(\ref{eq15}) functions $k_{\alpha}  (x)$
can have a complicated form. This would entail the expansion of the
integration region $G$ to observe the boundary condition (\ref{eq14}). For $n>20$ and
large $x$ Hermite's polynomials $H_{2n}(x)$ used in the DAF-based method can
acquire values that require the application of arbitrary-precision
arithmetic. This leads to even greater computational complexity and also
limits the application of the method.

Method (\ref{eq18}) is free from the above features. The necessity of choosing a
sufficiently dense uniform mesh can be classified as a limitation of scheme
(\ref{eq18}). However this can be avoided by choosing a mesh with a variable space
step. It is possible to construct an unconditionally stable homogeneous
conservative finite-difference scheme for problem (\ref{eq13}) - (\ref{eq15}) on a
non-uniform mesh using Refs. \cite{Samar39,Samar40}.

\section{SPATIALLY EXTENDED STOCHASTIC BRUSSELATOR}

\subsection {Mean-field result}

Let us apply the method developed in Sec. III to the study of probability
density of the system describing the well known model of autocatalytic
chemical reaction (brusselator \cite{Prig28}) with spatially correlated
multiplicative noise. Simultaneously let us study the variance of some
statistical first- and second-order characteristics of the system by
increasing the intensity of the external noise. In this paper the range of
parameters at which the Turing bifurcation arises in a deterministic system
is considered.

Brusselator is a model of a simple autocatalytic chemical reaction having a
trimolecular step \cite{Prig28}. The concentrations of the initial and final products
in this reaction are maintained constant. The influence of external
fluctuating environment can lead to the fact that concentrations of the
initial and final products become random functions. This leads to the
necessity of including noise into the kinetic equations of a deterministic
model. Let us assume that the concentration of the initial product
$B_{in}$ is most affected by external random environment. Then kinetic
equations of the reaction under consideration have the form:

\begin{equation}
\label{eq23}
{\frac{{\partial x_{1}}} {{\partial t}}} = A + x_{1} ^{2}x_{2} - (B + 1 +
\xi _{1} (\mathbf{r},t))x_{1} + D_{1} \nabla ^{2}x_{1} ,\\
{\frac{{\partial x_{2}}} {{\partial t}}} = - x_{1} ^{2}x_{2} + (B + \xi _{2}
(\mathbf{r},t))x_{1} + D_{2} \nabla ^{2}x_{2} ,
\end{equation}

\noindent
where $x_{1}$, $x_{2}$ are concentrations of intermediate components, $D_{1}$, $D_{2}$ are their diffusion coefficients, $A$, $B_{in}$ are concentrations of initial products with $B_{in} = B + \xi _{i}(\mathbf{r},t)$. Parameter $B$ is the spatio-temporal average of the initial product concentration $B_{in}$. The decrease in concentration $x_{1}$ is due to the two decays: with the formation of one of the final products and with the formation of an intermediate product $x_{2}$  and the second final product. These decays have different chemical reaction rates which are affected by external noises in different ways. It is taken into account by including different uncorrelated fields $\xi _{i} (\mathbf{r},t)$ into Eqs. (\ref{eq23}). Statistical properties of the fields $\xi _{i} (\mathbf{r},t)$  are described in Sec. I.

The system of equations (\ref{eq23}) is a specific case of Eq. (\ref{eq1}) with $n$ = 2. Therefore, multidimensional single-site NSCFPE in the Stratonovitch interpretation can be immediately written for the model (\ref{eq23}) using Eq. (\ref{eq12}):

\[
{\frac{{\partial w(x_{1} ,x_{2} ,t)}}{{\partial t}}} = {\frac{{\partial
}}{{\partial x_{1}}} }{\left\{ {{\left[ { - A - x_{1} ^{2}x_{2} + (B + 1 +
\theta _{1} )x_{1} - D_{1} (E(x_{1} \vert x_{2} ) - x_{1} )} \right]}w +
\theta _{1} \Phi _{1,0} x_{1}^{2} {\frac{{\partial w}}{{\partial x_{1}}} }}
\right\}} +
\]
\begin{equation}
\label{eq24}
{\frac{{\partial}} {{\partial x_{2}}} }{\left\{ {{\left[ {x_{1} ^{2}x_{2}
- Bx_{1} - D_{2} (E(x_{2} \vert x_{1} ) - x_{2} )} \right]}w + \theta _{2}
\Phi _{2,0} x_{1}^{2} {\frac{{\partial w}}{{\partial x_{2}}} }} \right\}},
\end{equation}

\[
E(x_{1} \vert x_{2} ,t) = {\int\limits_{ - \infty} ^{ + \infty}  {x_{1}
w(x_{1} \vert x_{2} ,t)dx_{1}}}  ,
\quad
E(x_{2} \vert x_{1} ,t) = {\int\limits_{ - \infty} ^{ + \infty}  {x_{2}
w(x_{2} \vert x_{1} ,t)dx_{2}}}  ,
\]

\[
w(x_{1} \vert x_{2} ,t) = {\frac{{w(x_{1} ,x_{2} ,t)}}{{{\int\limits_{ -
\infty} ^{ + \infty}  {w(x_{1} ,x_{2} ,t)dx_{1}}} } }},
\quad
w(x_{2} \vert x_{1} ,t) = {\frac{{w(x_{1} ,x_{2} ,t)}}{{{\int\limits_{ -
\infty} ^{ + \infty}  {w(x_{1} ,x_{2} ,t)dx_{2}}} } }}.
\]

Numerical solutions for Eq. (\ref{eq24}) are obtained using the finite-difference scheme (\ref{eq18}) (see Appendix A). Eq. (\ref{eq24}) has a greater computational complexity than, for example, the problems (\ref{eq19}) and (\ref{eq22}). This is related to the fact that, first, the drift coefficients $r_{1,2}(x_{1}, x_{2},t)$  are nonlinear and alternating-sign. This leads to the fact that regions with the different direction of the probability density drift arise on the plane $\left( {x_{1}, x_{2}}\right)$  (see Fig. 12). Moreover, the boundaries of these regions defined by the equations $r_{1,2}(x_{1}, x_{2},t) = 0$  are moving since conditional means $E(x_{1}\vert x_{2},t), E(x_{2}\vert x_{1},t)$ are time functions. Second, the diffusion coefficients are proportional to $x_{1}^{2}$, which leads to a significant increase of the integration region necessary to satisfy the boundary conditions (\ref{eq14}). Third, the problem (\ref{eq24}) is two-dimensional and increasing the dimension of the space always leads to an increase of computational complexity.

\begin{figure}[!h]
\centering
\includegraphics[width=2.50in,height=2.04in]{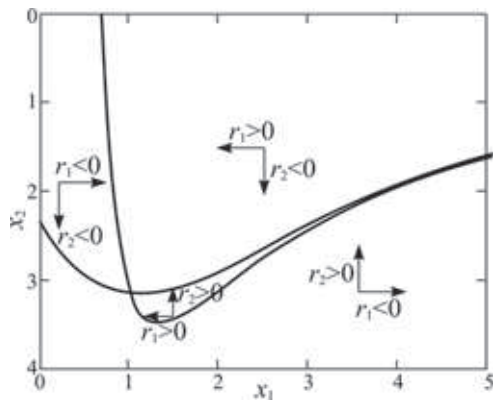}
\caption{Probability density (\ref{eq24}) drift directions on the plane $\left( {x_{1}, x_{2}}\right)$. Parameters are $A$=3, $B$=7, $\theta _{1} = \theta _{2} = 0.1$,
$D_{1}$=1, $D_{2}$=5, $E(x_{1} \vert x_{2} ,t) = A, E(x_{2} \vert
x_{1} ,t) = B / A$, $\Phi _{1,0} = \Phi _{2,0} = 1$.}
\end{figure}

Figures 13, 17, 21, 22 present characteristic types of solutions (\ref{eq24}) obtained at different values of the parameters of the problem and the noise intensity. The initial distribution is Gaussian with variances $\theta_{1}$  and $\theta_{2}$, and expectations equal to stationary values of $x_{10}$ and $x_{20}$ in the absence of noise (see Appendix A). The following parameters for numerical integration (\ref{eq24}) remain constant in our calculations: $A$=3, $D_{1}=1$, $D_{2}=5$, $\Phi
_{1,0} = \Phi _{2,0} = 1$. The other parameters are indicated under the figures. The critical value of parameter $B_{ñ}$ is 5.47 in a deterministic case at given $A$, $D_{1}$, and $D_{2}$.

Fig. 13 demonstrates the evolution of probability density  $w(x_{1} ,x_{2}
,t)$ in the vicinity of the deterministic bifurcation point and small noise intensity. It can be seen from this figure that the symmetry of initial distribution is violated in the evolution process. The probability density distribution remains unimodal throughout the time until the stationary state is reached. Hence the state of the system (\ref{eq23}) is ordered, despite the noise.

\begin{figure}
\centering
\includegraphics[width=3.08in,height=6.40in]{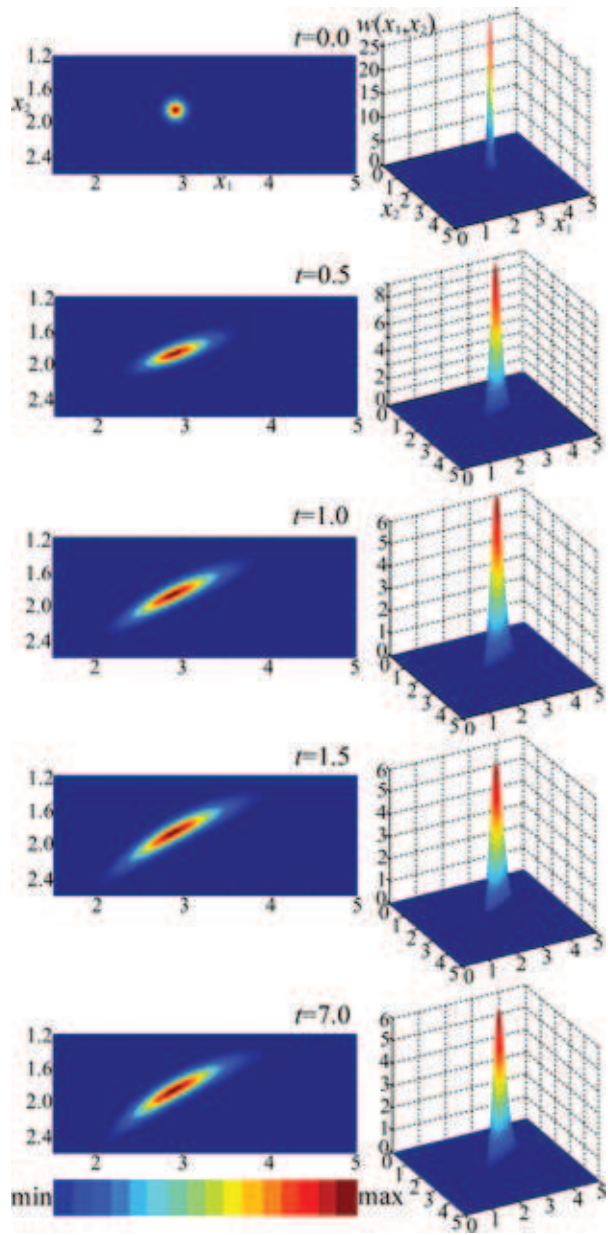}
\caption{The evolution of probability density (\ref{eq24}) for the model (\ref{eq23}). Unimodal distribution (top view in the left-hand side). The color gradient from dark blue to dark red visualizes the change from minimum to maximum. The model parameters are $B$=5.5, $\theta _{1} = \theta _{2} = 0.005$. The time moment $t$=7 corresponds to the stationary state.}
\end{figure}

\begin{figure}
\centering
\includegraphics[width=2.73in,height=1.55in]{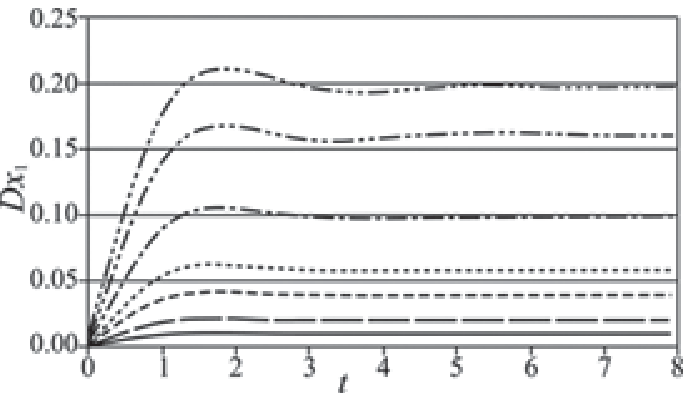}\\
a)\\
\includegraphics[width=2.93in,height=1.66in]{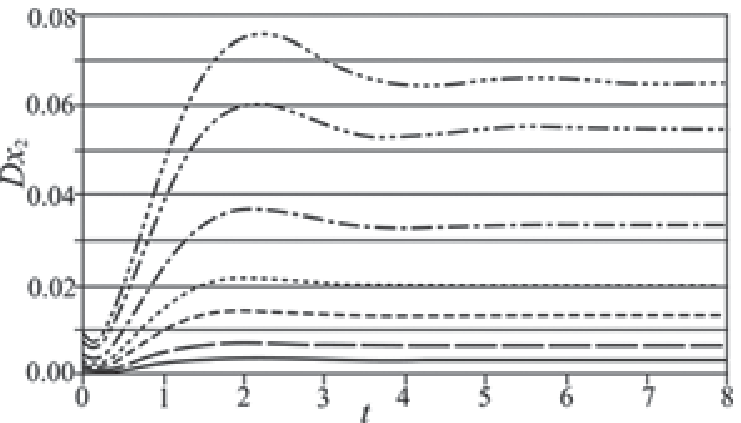}\\
b)\\
\caption{The dependencies of variance $Dx$ of concentration on time with increasing noise intensity:
a) first product, b) second product. The solid line $\theta _{1} = \theta _{2} = \theta = 0.0005$, the line with a long dash $\theta = 0.001$, the dashed line $\theta = 0.002$, the dotted line $\theta = 0.003$, the dash-dotted line $\theta = 0.005$, the dash-dot-dot line $\theta = 0.008$, the three dots - dash line $\theta = 0.01$. $B$=5.5}
\end{figure}

\begin{figure}
\centering
\includegraphics[width=3.21in,height=2.11in]{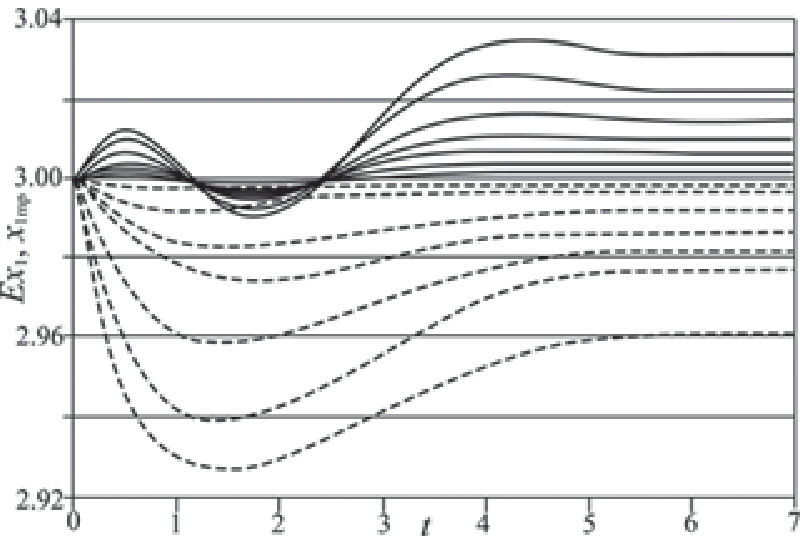}\\
a)\\
\includegraphics[width=3.25in,height=1.95in]{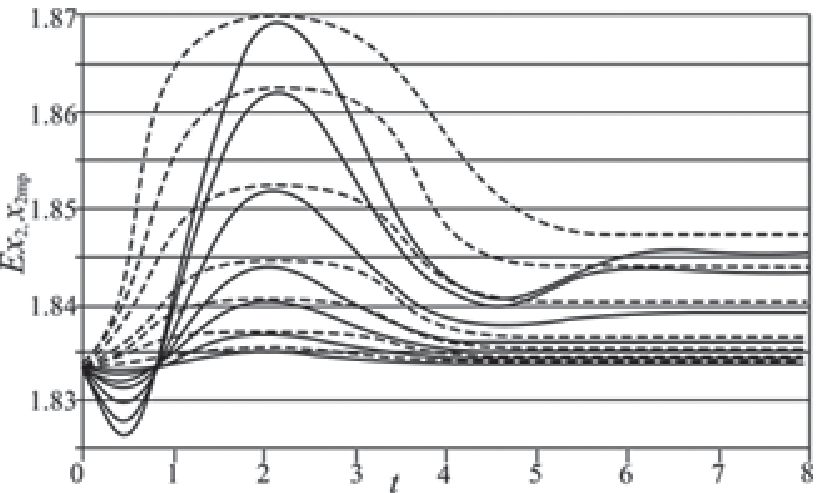}\\
b)\\
\caption{Fig. 15. The dependencies of mean $Ex$ (solid lines) and most probable $x_{\rm {mp}}$ (dashed lines) values on time in case of increasing noise intensity: a) first product, b) second product. $B$=5.5. $\theta _{1}, \theta_{2}$ are as in Fig. 14. The greater the noise intensity, the greater the deviation of values $Ex$ and $x_{\rm {mp}}$ from the stationary values of $x_{10}$ and $x_{20}$ ($t$=0).}
\end{figure}

\begin{figure}
\centering
\includegraphics[width=3.19in,height=2.05in]{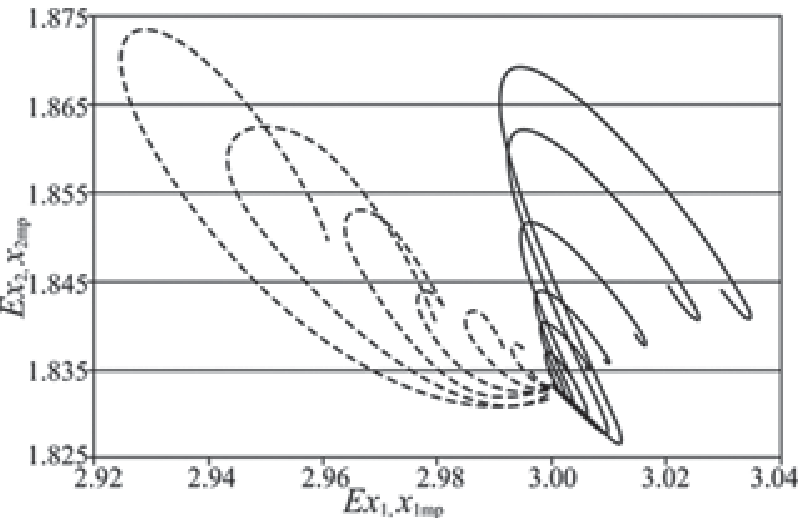}\\
\caption{Changes of the mean $Ex$ (solid lines) and most probable $x_{\rm {mp}} $ (dashed lines) values $x_{1}$ and $x_{2}$ in case of increasing noise intensity for unimodal distribution. $B$=5.5. $\theta _{1}
, \theta _{2} $  are as in Fig. 14. The greater the noise intensity, the greater the size of the wreath of the curve.}
\end{figure}

Fig. 14 shows the appropriate dependencies of variance of concentrations $x_{1}$ and $x_{2}$ on time if noise intensity increases. It can be seen that the greater external noise intensity, the faster the variance increases and the greater its value in the stationary state. Fig. 15 demonstrates the dependencies of the mean and most probable values on time with different noise intensities. The increase of noise intensity leads to the increase of difference between the mean and the appropriate most probable in the steady stationary state. Fig. 16 illustrates this more clearly. All the results given above are quite expectable.

Quite a different picture is observed at a greater distance from the deterministic bifurcation point. Fig. 17 presents a more complicated type of the probability density $w(x_{1},x_{2},t)$  evolution. We can see that at first unimodal distribution is conserved, but it is strongly "blurred" and the maximum of density drifts away from the initial position. Then the splitting of probability density with the formation of two maxima takes place at a certain point in time. Bimodal probability distribution "lives" during a certain time interval. Herewith competition occurs between maxima, as a result of which one of the maxima suppresses the other. The distribution becomes unimodal again. That is, transient bimodality is observed in the ordered phase. The behavior of the statistical characteristics also varies as the noise intensity increases.

Fig. 18 shows the dependencies of variance $Dx_{1}$ of concentration of the first product on time when the noise intensity increases. Figs. 19, 20 demonstrate the appropriate changes of the mean and most probable values. Dependencies $Dx_{1}(t)$,  $Ex_{i}(t), x_{i{\rm mp}}(t)$ are similar to the ones given in Figs. 14,15,16 if the noise intensities $\theta _{1}, \theta _{2} < 0.09$. The distribution remains unimodal. A clearly visible "dip" corresponding to the disappearance of transient bimodality is observed in the dependence $Dx_{1}(t)$ at $\theta _{1}, \theta _{2} \ge 0.09$. Herewith the discontinuity of the first kind appears in plots of the most probable vs time. The jump from a wreath of the curve $x_{2{\rm mp}}(x_{1{\rm mp}})$ corresponds to the disappearance of transient bimodality in Fig. 20.

\begin{figure}
\centering
\includegraphics[width=3.15in,height=6.52in]{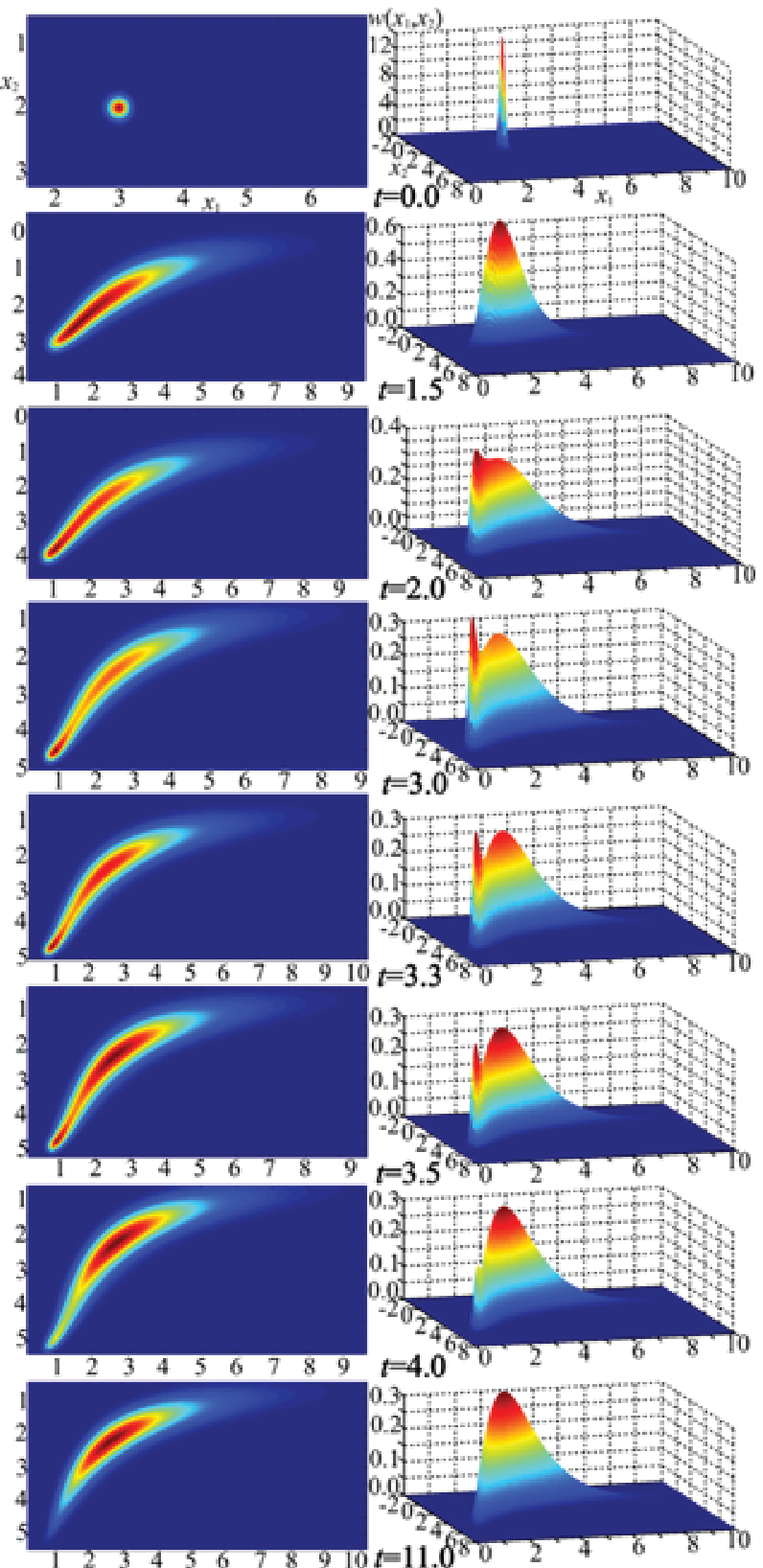}\\
\caption{The evolution of probability density (\ref{eq24}) for the model (\ref{eq23}). Transient bimodality is observed in the time interval $t \in [1.5,4]$. Model parameters are $B$=6, $\theta _{1} = \theta _{2} = 0.09$. The time moment $t$=11 corresponds to the setting of the stationary state.}
\end{figure}

\begin{figure}
\centering
\includegraphics[width=3.20in,height=2.08in]{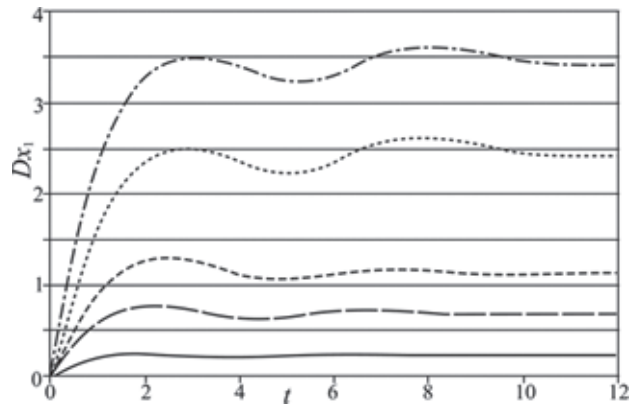}\\
\caption{Dependencies of variance $Dx_{1}$ of concentration $x_{1}$ on time with increasing noise intensity. The solid line $\theta_{1} = \theta_{2} = \theta = 0.01$, the line with a long dash $\theta = 0.03$, the dashed line $\theta = 0.05$, the dotted line $\theta = 0.09$, the dash-dotted line $\theta = 0.12$. $B$=6.}
\end{figure}

\begin{figure}[!h]
\centering
\includegraphics[width=3.12in,height=1.69in]{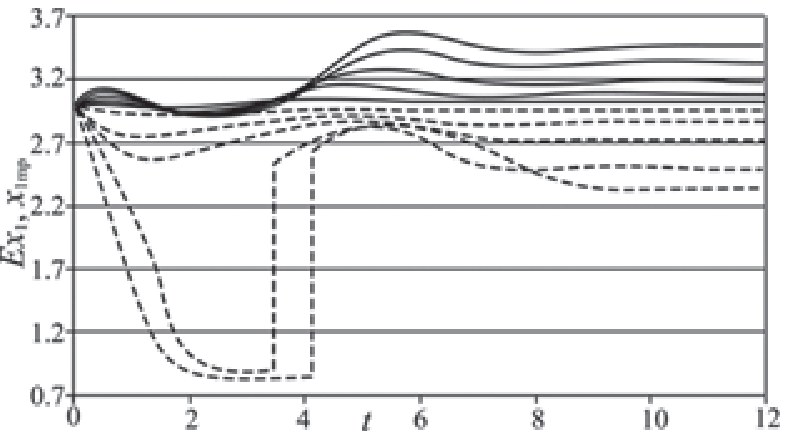}\\
a)\\
\includegraphics[width=3.01in,height=1.66in]{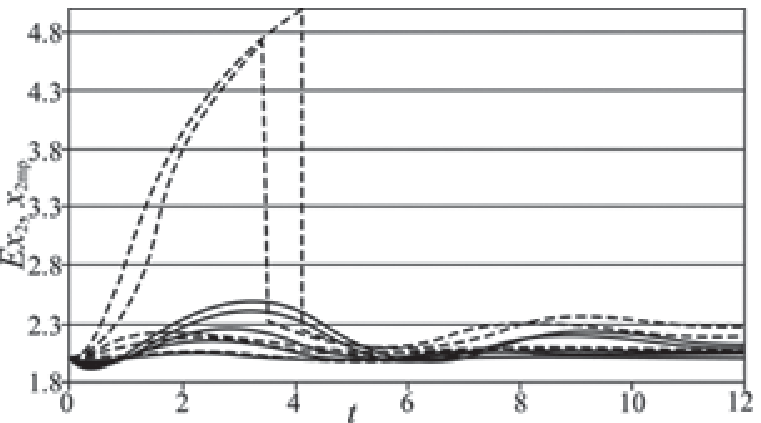}\\
b)\\
\caption{The dependencies of mean $Ex$ (solid lines) and most probable $x_{\rm{mp}}$ (dashed lines) values on time with increasing noise intensity: a) first product, b) second product. $B$=6. $\theta _{1}, \theta _{2}$ are as in Fig. 14. The greater the noise intensity, the greater the deviation of values $Ex$ and $x_{\rm{mp}}$  from the stationary values of $x_{10}$ and $x_{20}$. The most probable jump (discontinuity of the first kind) corresponds to the disappearance of bimodality.}
\end{figure}

\begin{figure}[!h]
\centering
\includegraphics[width=3.24in,height=1.93in]{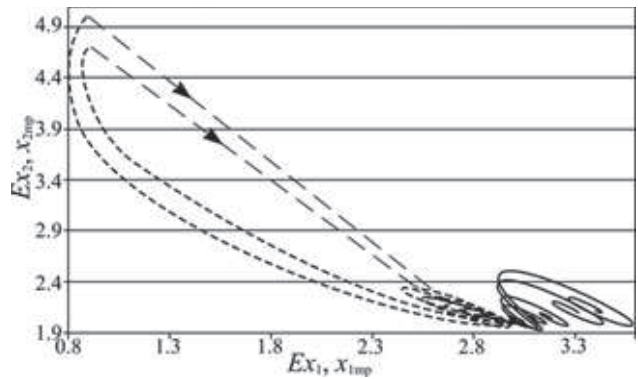}\\
\caption{Changes of the mean $Ex$ (solid lines) and the most probable $x_{\rm{mp}}$ (dashed lines) values $x_{1}$ and $x_{2}$ in case of increasing the noise intensity. $B$=6, the other model parameters are as in Fig. 18. A jump from a wreath of the curve $x_{2{\rm mp}}(x_{1{\rm mp}})$ shown in the figure by a thin line with a dash and the arrow corresponds to the disappearance of bimodality.}
\end{figure}

\begin{figure}[!h]
\centering
\includegraphics[width=4.50in,height=6.73in]{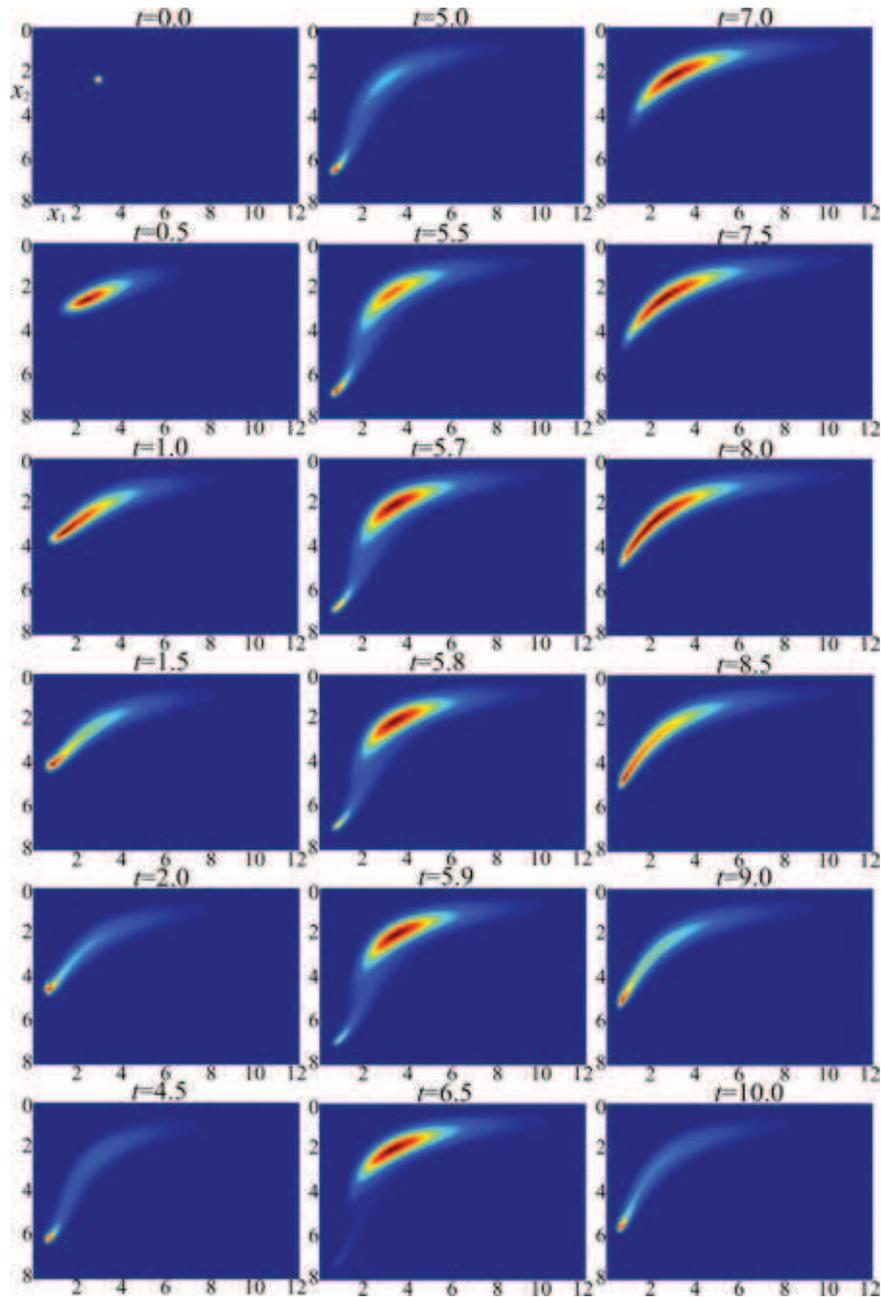}\\
\caption{The evolution of probability density (\ref{eq24}) for the model (\ref{eq23}). "Repumping" of probability density through bimodality (top view). The model parameters are $B$=7, $\theta_{1} = \theta_{2} = \theta = 0.1$. The figure presents one "period" of "repumping". The sequences of the frames in the left-hand side and the right-hand side correspond to unimodal distribution, while in the center they correspond to the bimodal one.}
\end{figure}

\begin{figure}[!h]
\centering
\includegraphics[width=3.00in,height=6.56in]{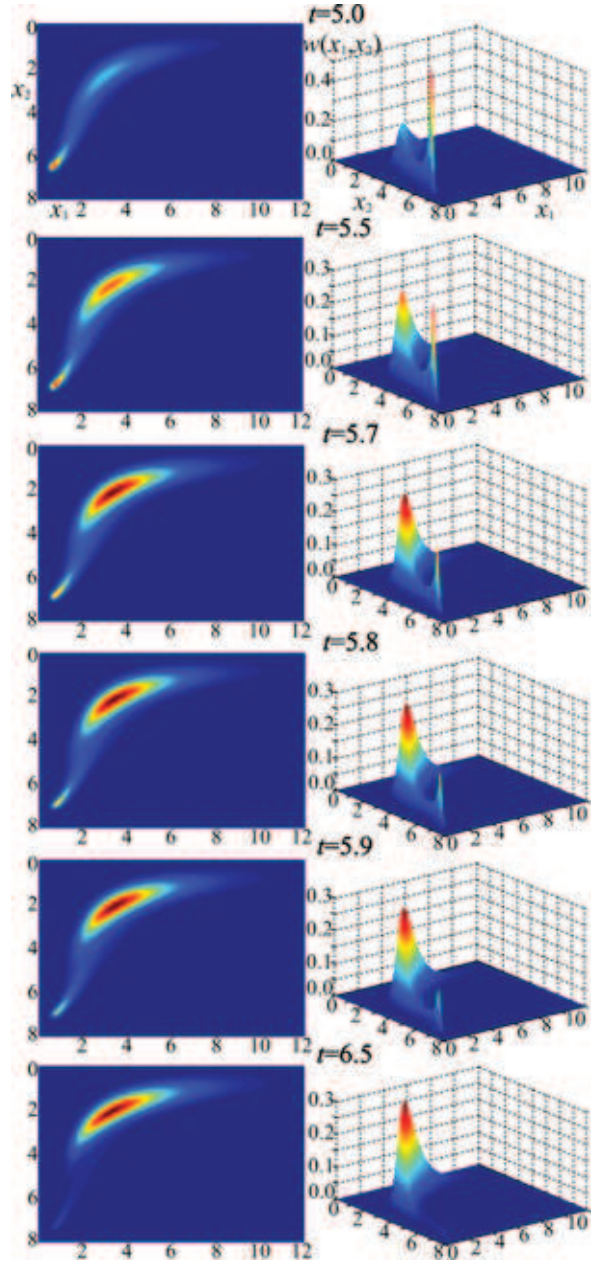}\\
\caption{"Repumping" of probability density through bimodality (corresponds to the center of Fig.21).}
\end{figure}

\begin{figure}[!h]
\centering
\includegraphics[width=2.95in,height=1.68in]{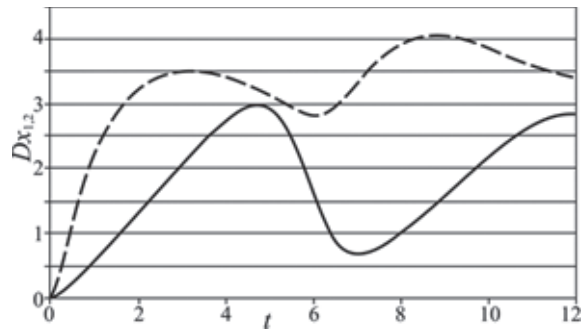}\\
\caption{Dependencies of variances $Dx_{1}$ and $Dx_{2}$ on time at $B$=7, $\theta_{1} = \theta_{2} = \theta = 0.1$. The figure presents one "period of repumping".}
\end{figure}

\begin{figure}[!h]
\centering
\includegraphics[width=3.15in,height=2.00in]{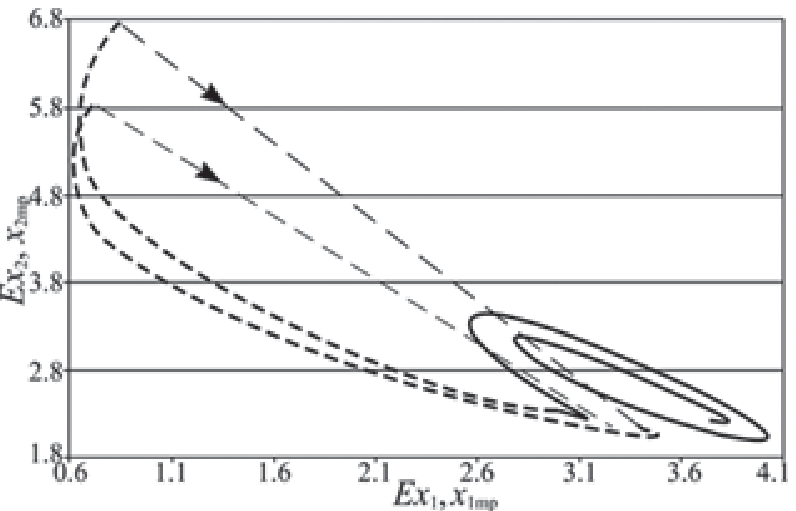}\\
\caption{Change of the mean (solid line) and the most probable (dashed line) values $x_{1}$ and $x_{2}$ in "repumping". Two "periods" are presented.}
\end{figure}

A completely unexpected solution (\ref{eq24}) appears at a greater distance from the deterministic point of bifurcation (see Figs. 21-24). At first density drifts from the initial position to the boundary of the integration region in accordance with the directions indicated in Fig. 12. Then the splitting of density occurs at $t\sim $5 (Figs. 21, 22) just as in transient bimodality.

Peculiar "repumping" of probability density from one maximum to another  through bimodality is observed until the time moment $t\sim $6.5 (Fig. 22). It can be noticed that the duration of the existence of one- and bimodal distributions are comparable in the order of magnitude. Then a drift towards the boundaries happens again. The process is repeated until the stationary state is established. This is accompanied by a gradual decrease of maximum values in $Dx_{i}(t)$ (Fig. 23), and decrease in the size of the wreath of the curve $Ex_{2}(Ex_{1})$ (Fig. 24). Figure 24 shows two jumps from a wreath of the curve $x_{2\rm{mp}}(x_{1\rm{mp}})$, which corresponds to the double appearance and disappearance of "repumping" of the probability density through bimodality.

Such behavior of the probability density implies multiple appearance of the other state (other phase) that corresponds to bimodal distribution, in the ordered phase. We can assume that there is a kind of phase "intermittency". This noise-induced effect will be presented in more detail in our future paper.

So, as a result of the numerical study of Eq. (\ref{eq24}) solutions we found that different type of solutions can arise in the region of Turing bifurcation when noise intensity increases: unimodal distribution, unimodal distribution with transient bimodality, and complicated distribution, in which unimodal and bimodal distributions alternate until the steady state is established. In other words, only the ordered phase is observed at low noise intensity. The increase of the noise intensity leads to the appearance of transient bimodality (disordered phase) in the ordered phase. Further growth of the noise intensity disrupts the ordering to an even greater extent: there is an "intermittency" phase, which "swings" the ordered state, as it were.

\subsection{Fokker-Planck equation for order parameters}

The system (\ref{eq23}) analysis presented above takes into account the interaction of the whole set of both stable and unstable modes. It is known that the system behavior is governed by the behavior of unstable modes (order parameters) \cite{Haken41} in the vicinity of the Turing bifurcation point. Therefore, additionally we study the behavior of order parameters of this system. The procedure of deriving generalized Ginzburg - Landau equations for type (\ref{eq1}) systems was proposed in Ref. \cite{Kur27}. Following this procedure we obtained stochastic equations for the amplitudes of unstable modes for system (\ref{eq23}). These equations have the form:

\begin{equation}
\label{eq25}
{\frac{{d\xi _{\mathbf{k} u}^{(1)}}} {{d\tau}} } = F_{\mathbf{k} u}(\tau ),
\end{equation}

\[
F_{\mathbf{k} u} (\tau ) = \lambda _{1} ({\mathbf{k}}_{u} )\xi _{\mathbf{k} u}^{(1)} +
{\sum\limits_{{\mathbf{k}}' u} {\Omega _{1} ({\mathbf{k}}_{u} ,{\mathbf{k}'}_{u} ,{\mathbf{k}}_{s} ,{\mathbf{z}}(\tau ))\xi
_{{\mathbf{k}}'u}^{(1)}}}   +
\]
\[
{\sum\limits_{{\mathbf{k}}'u{\mathbf{k}}''u} {\Omega _{11} ({\mathbf{k}}_{u} ,{\mathbf{k}'}_{u}, {\mathbf{k}''}_{u}, {\mathbf{k}}_{s}, {\mathbf{z}}(\tau ))}} \xi _{{\mathbf{k}}'u}^{(1)} \xi _{{\mathbf{k}}''u}^{(1)} +
{\sum\limits_{{\mathbf{k}}'u{\mathbf{k}}''u{\mathbf{k}}'''u} {\omega ({\mathbf{k}}_{u}, {\mathbf{k}'}_{u} ,{\mathbf{k}''}_{u}, {\mathbf{k}'''}_{u} )}} \xi _{{\mathbf{k}}'u}^{(1)} \xi _{{\mathbf{k}}''u}^{(1)} \xi _{{\mathbf{k}}'''u}^{(1)} +
\]
\[
A{\left[ { - O_{1}^{ * (1)} ({\mathbf{k}}_{u})z_{1,{\mathbf{k}}u} (\tau )
+ O_{2}^{ * (1)} ({\mathbf{k}}_{u} )z_{2,{\mathbf{k}}u} (\tau )} \right]}-
{\sum\limits_{\mu,\varphi,{\varphi}' = 1}^{2} {{\sum\limits_{{\mathbf{k}}s} {\zeta _{\varphi {\varphi}'}^{(\mu )} ({\mathbf{k}}_{s},{\mathbf{k}}_{u} )z_{{\varphi} ',{\mathbf{k}}u - {\mathbf{k}}s} (\tau )}}}
}z_{\varphi ,{\mathbf{k}}s} (\tau ).
\]

Here $\xi _{{\mathbf{k}}u}^{(1)}$ are unstable mode amplitudes of system
(\ref{eq23}), ${\mathbf{k}}_{u} , {\mathbf{k}}_{s} $ are wave numbers of unstable
and stable modes respectively, ${\mathbf{z}}(\tau )$ is the random vector
field, the components of which $z_{\varphi,{\mathbf{k}}} (\tau ) = \int
{\xi _{\varphi}  ({\mathbf{r}},\tau )e^{- i{\mathbf{k}}{\mathbf{r}}}d{\mathbf{r}}} $
have zero means and a given correlation tensor
$K[z_{j,{\mathbf{k}}}(t), z_{l,{\mathbf{k}'}} (\tau )] = g_{jl} ({\left|
{{\mathbf{k}}} \right|})\delta ({\mathbf{k}} - {\mathbf{k}'})\delta (t
- \tau )\delta _{jl} $,  $\varphi$ and $\mathbf{k}$ are index arguments of this field.
Taking into account that the functions $\Phi _{i}({\left| {{\mathbf{r}} -
{\mathbf{r}'}} \right|})$ in Eqs. (\ref{eq1}) were chosen to be exponential for
definiteness, for two-dimensional media $g_{ii} = \theta _{i} k_{fi} / [2\pi
^{2}(k^{2} + k_{fi}^{2} )^{ - 3 / 2}]$. Functions $\lambda _{1} ({\mathbf{k}}_{u} )$, $\Omega _{1} ({\mathbf{k}}_{u},
{\mathbf{k}'}_{u}, {\mathbf{k}}_{s}, {\mathbf{z}}(\tau ))$, $\Omega _{11} ({\mathbf{k}}_{u} ,{\mathbf{k}'}_{u}, {\mathbf{k}''}_{u}, {\mathbf{k}}_{s}, {\mathbf{z}}(\tau )$
and others introduced in Eq. (\ref{eq25}) are presented in Appendix B.

Equations (\ref{eq25}) define the evolution of a set of random processes. Let us write FPE for these processes. It can be represented in a general form as follows:

\begin{equation}
\label{eq26}
{\frac{{\partial w\left( {{\left\{ {\xi _{{\mathbf{k}}u}^{(1)}}
\right\}},\tau}  \right)}}{{\partial \tau}} } = - {\sum\limits_{{\mathbf{k}}u} {{\frac{{\partial}} {{\partial \xi _{{\mathbf{k}}u}^{(1)}}} }{\left\{
{\left( { < F_{{\mathbf{k}}u} (\tau ) > + {\sum\limits_{{\mathbf{q}}u}
{{\int\limits_{ - \infty} ^{0} {K[{\frac{{\partial F_{{\mathbf{k}}u} (\tau
)}}{{\partial \xi _{{\mathbf{q}}u}^{(1)}}} },F_{{\mathbf{q}}u} ({t}')]}}}
}d{t}'} \right)w} \right\}}}}  +
\end{equation}

\[
 + {\sum\limits_{{\mathbf{k}}u,{\mathbf{q}}u} {{\frac{{\partial
^{2}}}{{\partial \xi _{{\mathbf{k}}u}^{(1)} \partial \xi _{{\mathbf{q}}u}^{(1)}}} }{\left\{ {\left( {{\int\limits_{ - \infty} ^{0} {K[F_{{\mathbf{k}}u} (\tau ),F_{{\mathbf{q}}u} ({t}')]}} d{t}'} \right)w} \right\}}}
}.
\]

\noindent
Here $w\left( {{\left\{ {\xi _{{\mathbf{k}}u}^{(1)}}  \right\}},\tau} \right)$  is the multivariate probability distribution density defining the probability of some configuration of unstable modes ${\left\{ {\xi _{{\mathbf{k}}u}^{(1)}}  \right\}}$ . After transformations with an accuracy up to the terms linear in the noise intensity one can obtain the correlation functions appearing in Eq. (\ref{eq26}) that are presented in Appendix C.

Let the space of the system under study be two-dimensional. If only one mode with the wave number ${\mathbf{k}}_{c} $ and amplitude $\xi _{{\mathbf{k}}c} $ is unstable in such space the Eq. (\ref{eq26}) acquires a simple structure:

\begin{equation}
\label{eq27}
{\frac{{\partial w(\xi _{{\mathbf{k}}c} ,\tau )}}{{\partial \tau}} } = -
{\frac{{\partial}} {{\partial \xi _{{\mathbf{k}}c}}} }{\left\{ {(h + a\xi_{{\mathbf{k}}c} + b\xi _{{\mathbf{k}}c}^{3} )w - (c + d\xi _{{\mathbf{k}}c}^{2} + e\xi _{{\mathbf{k}}c}^{4} ){\frac{{\partial w}}{{\partial \xi
_{{\mathbf{k}}c}}} }} \right\}}.
\end{equation}

Constants $a$, $b$, $c$, $d$, $e$, $h$ are easily obtained assuming that ${\mathbf{k}'}_{u}
= {\mathbf{q}'}_{u} = {\mathbf{q}''}_{u} = {\mathbf{k}}_{c} $ in
correlators $K[{\frac{{\partial F_{{\mathbf{k}}u} (\tau )}}{{\partial \xi
_{{\mathbf{q}}u}^{(1)}}} },F_{{\mathbf{q}}u} ({t}')]$, $K[F_{{\mathbf{k}}u} (\tau ),F_{{\mathbf{q}}u} ({t}')]$ (see Appendix C).

The stationary solution of Eq. (\ref{eq27}) has the form:

\begin{equation}
\label{eq28}
w_{st} (\xi _{{\mathbf{k}}c} ) = {\left\{ {{\begin{array}{*{20}c}
 {N{\left| {c + d\xi _{{\mathbf{k}}c}^{2} + e\xi _{{\mathbf{k}}c}^{4}}
\right|}^{{\frac{{b}}{{4e}}}}{\left| {{\frac{{2e\xi _{{\mathbf{k}}c}^{2} +
d - \sqrt {d^{2} - 4ec}}} {{2e\xi _{{\mathbf{k}}c}^{2} + d + \sqrt {d^{2} -
4ec}}} }} \right|}^{{\frac{{2ae - bd}}{{4e\sqrt {d^{2} - 4ec}}} }}\exp I,    d^{2} > 4ec,} \hfill \\
{N{\left| {c + d\xi _{{\mathbf{k}}c}^{2} + e\xi _{{\mathbf{k}}c}^{4}}
\right|}^{{\frac{{b}}{{4e}}}}\exp {\left\{ {{\frac{{2ae - bd}}{{2e\sqrt {4ec
- d^{2}}}} } \arctan{\left( {{\frac{{2e\xi _{{\mathbf{k}}c}^{2} + d}}{{\sqrt {4ec - d^{2}}}} }} \right)}} \right\}} \exp
I,   4ec > d^{2}.} \hfill \\
\end{array}}}  \right.}
\end{equation}

Here

\[
I = {\left\{ {{\begin{array}{*{20}c}
 {{\frac{{eh}}{{\sqrt {d^{2} - 4ec}}} }(I_{1} - I_{2} ), d^{2} > 4ec,} \hfill \\
 {{\frac{{h}}{{4e\sin \alpha}} }{\left[ {\sin {\frac{{\alpha}} {{2}}}\ln
\left( {{\frac{{\xi _{{\mathbf{k}}c}^{2} + 2q\xi _{{\mathbf{k}}c} \cos
{\frac{{\alpha}} {{2}}} + q^{2}}}{{\xi _{{\mathbf{k}}c}^{2} - 2q\xi _{{\mathbf{k}}c} \cos {\frac{{\alpha}} {{2}}} + q^{2}}}}} \right) + 2\cos
{\frac{{\alpha}} {{2}}} \arctan {\left(
{{\frac{{\xi _{{\mathbf{k}}c}^{2} - q^{2}}}{{2q\xi _{{\mathbf{k}}c}
\sin {\frac{{\alpha}} {{2}}}}}}} \right)}} \right]},	4ec > d^{2}.} \hfill \\
\end{array}}}  \right.}
\]

\[
\begin{array}{l}
 \cos \alpha = - d / (2\sqrt {ec} ),	q = \sqrt[{4}]{{c / e}},	f_{1,2} = d /
2 \mp (d^{2} - 4ec)^{1 / 2} / 2. \\
 I_{1,2} = {\left\{ {{\begin{array}{*{20}c}
 {{\frac{{1}}{{\sqrt {ef_{1,2}}} } }\arctan \left( {\xi _{{\mathbf{k}}c}^{} \sqrt {{\frac{{e}}{{f_{1,2}}} }}} \right), ef_{1,2} > 0,} \hfill \\
 {{\frac{{1}}{{2i\sqrt {ef_{1,2}}} } } \ln \left( {{\frac{{f_{1,2}
+ i\xi _{{\mathbf{k}}c}^{} \sqrt {ef_{1,2}}} } {{f_{1,2} - i\xi _{{\mathbf{k}}c}^{} f_{1,2}}} }} \right),ef_{1,2} < 0.} \hfill \\
\end{array}}}  \right.} \\
 \\
 \end{array}
\]

$N$ is the normalization constant:

\[
N = {\left\{ {{\begin{array}{*{20}c}
 {1 / {\int\limits_{ - \infty} ^{ + \infty}  {\exp I{\left| {c + d\xi _{{\mathbf{k}}c}^{2} + e\xi _{{\mathbf{k}}c}^{4}}
\right|}^{{\frac{{b}}{{4e}}}}{\left| {{\frac{{2e\xi _{{\mathbf{k}}c}^{2} +
d - \sqrt {d^{2} - 4ec}}} {{2e\xi _{{\mathbf{k}}c}^{2} + d + \sqrt {d^{2} -
4ec}}} }} \right|}^{{\frac{{2ae - bd}}{{4e\sqrt {d^{2} - 4ec}}} }}d\xi
_{{\mathbf{k}}c}}},  d^{2} > 4ec,} \hfill \\
 {1 / {\int\limits_{ - \infty} ^{ + \infty}  {\exp I{\left| {c + d\xi _{{\mathbf{k}}c}^{2} + e\xi _{{\rm {\bf k}}c}^{4}}
\right|}^{{\frac{{b}}{{4e}}}}\exp {\left\{ {{\frac{{2ae - bd}}{{2e\sqrt {4ec
- d^{2}}}} }{\arctan}\left( {{\frac{{2e\xi _{{\mathbf{k}}c}^{2} + d}}{{\sqrt {4ec - d^{2}}}} }} \right)} \right\}}d\xi _{{\mathbf{k}}c}}}  ,		4ec > d^{2}.} \hfill \\
\end{array}}}  \right.}
\]

Fig. 25 demonstrates the steady-state probability density (\ref{eq28}) for the values of the critical mode amplitude of system (\ref{eq23}) in the supercritical region for different values of noise intensity. It can be seen from Fig. 25 that two maxima merge into one at $\theta _{1} = \theta _{2} = 5.5 \times
10^{ - 3} $(dashed line) and bimodal distribution is replaced by unimodal. The plot $w_{st}(\xi _{kc} )$  acquires a flat top (plateau). Herewith the steady-state most probable value of the critical mode amplitude module ${\left| {\xi _{{\mathbf{k}}c\,{\rm {mp}}}}  \right|}$  becomes zero (see Fig. 26). It also follows from Fig. 26 that as the distance from the bifurcation point increases, i.e. with the increase of the bifurcation parameter B, the noise intensity, at which ${\left| {\xi _{{\mathbf{k}}c\,{\rm {mp}}}}  \right|} = 0$ , increases. The steady-state mean ${\left\langle {{\left| {\xi _{{\mathbf{k}}c}}  \right|}} \right\rangle} $ is always other than zero and the difference between ${\left| {\xi _{{\mathbf{k}}c\,{\rm {mp}}}}  \right|} $  and ${\left\langle {{\left| {\xi _{{\mathbf{k}}c}}  \right|}} \right\rangle} $  increases both with the increase of the noise intensity and that of the parameter B. The latter corresponds to the conclusions from the plots presented in Figs. 16, 20.

Figure 27 illustrates the behavior of the steady-state second order cumulant
$\kappa_{2} =  {\left\langle{\xi_{\mathbf{k}c}^{2}}\right\rangle}/ {{ \left\langle |{\xi_{\mathbf{k}c}}| \right\rangle}^{2}} $
and the susceptibility
$ [ {\left\langle{\xi_{\mathbf{k}c}^{2}}\right\rangle} - {{ \left\langle |{\xi_{\mathbf{k}c}}| \right\rangle}^{2}} ]/{\theta}$
of the order parameter as the noise intensity increases at different values of the bifurcation parameter. The second order cumulant is a monotone increasing function at low noise whereas the susceptibility has a marked maximum. This maximum is observed for the values of  noise intensity slightly smaller than the values at which ${\left| {\xi _{{\mathbf{k}}c\,{\rm {mp}}}} \right|} = 0$ . This maximum can be called a "forerunner" of a change in the system state.

\begin{figure}[!h]
\centering
\includegraphics[width=2.94in,height=2.10in]{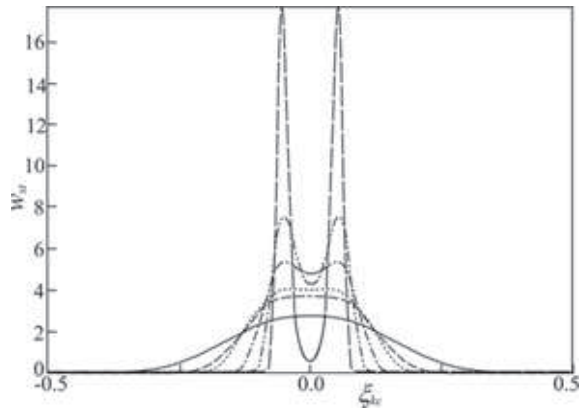}\\
\caption{Steady-state probability density, Eq. (\ref{eq28}), for the values of the critical mode amplitude of system (\ref{eq23}) in the supercritical region for six values of noise intensity. $B$=5.5. The line with a long dash $\theta _{1} =
\theta _{2} = \theta = 3.5 \times 10^{ - 5}$, the dash-dot-dot line $\theta =
2.0 \times 10^{ - 4}$, the dash-dot line $\theta = 8.0 \times 10^{ - 4}$, the dotted line $\theta = 3.0 \times 10^{ - 3}$, the dashed line $\theta = 5.5 \times 10^{ - 3}$, the solid line $\theta = 2.0 \times 10^{ - 2}$.}
\end{figure}

\begin{figure}[!h]
\centering
\includegraphics[width=2.85in,height=1.61in]{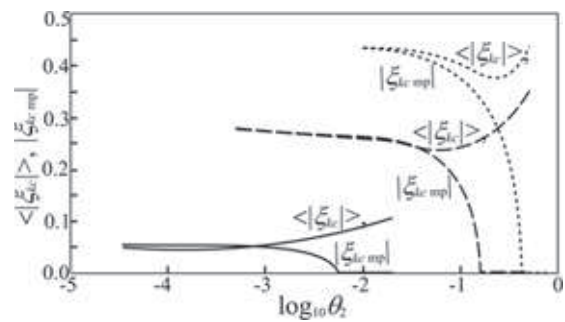}\\
\caption{Steady-state mean ${\left\langle {{\left| {\xi _{{\mathbf{k}}c}}
\right|}} \right\rangle} $ and most probable ${\left| {\xi _{{\mathbf{k}}c\,\rm {mp}}}  \right|}$ values of the critical mode amplitude
module as a function of noise intensity. The solid line $B$=5.5, the dashed
line $B$=6.0, the dotted line $B$=7.}
\end{figure}

\begin{figure}[!h]
\centering
\includegraphics[width=2.66in,height=1.58in]{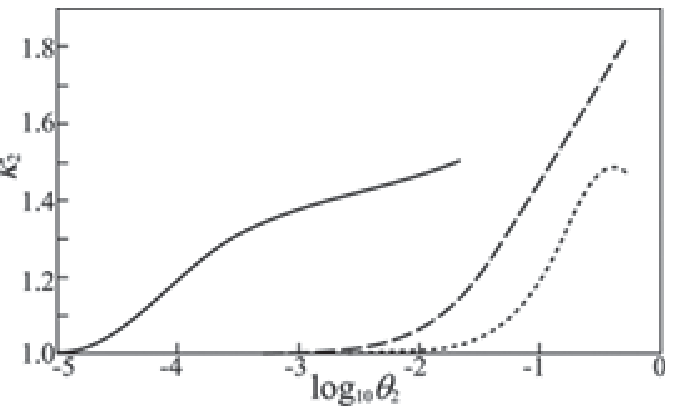}\\
a)\\
\includegraphics[width=2.63in,height=1.57in]{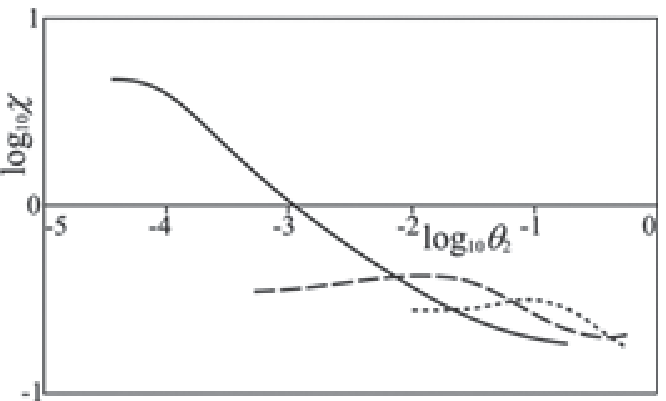}\\
b)\\
\caption{Steady-state second-order cumulant $\kappa _{2}$  (a) and susceptibility $\chi$   (b) as functions of noise intensity . $\theta _{2} = \theta _{1} = \theta$. The solid line $B=5.5$, the dashed line $B=6.0$, the dotted line $B=7$.}
\end{figure}

The analysis of the plots presented in Figs. 14, 18 shows that the
steady-state variance of system (\ref{eq23}) increases in the region of low noise.
This qualitatively corresponds to the areas where the susceptibility of the
order parameter in Fig. 27 (b) increases.

One can observe an interesting correspondence between the values of noise
intensity at which steady-state bimodal distribution disappears for the
critical mode amplitude of system (\ref{eq23}) and transient bimodality arises in
the ordered phase. We turn to figures 19 and 26. At $B$ = 6 these values are $
\approx 0.16$ and $ \approx 0.09 - 0.12$ respectively.

\section{CONCLUSION}

Mean field approximation was developed for studying the state of multicomponent stochastic spatially extended systems. We assume that in this case equality (\ref{eq10}) is true. Herewith the non-trivial spatial structure, spatial homogeneity, and isotropy of multiplicative noise are taken into account. In this approximation a multivariate single-site nonlinear self-consistent Fokker-Planck equation was derived for probability density of the state of the system under consideration.

The finite-difference method is proposed for the numerical solution of the general class of multivariate nonlinear self-consistent time-dependent Fokker-Planck equations. The accuracy and reliability of the method was illustrated on known one-dimensional problems. It was shown that the method proposed conserves the positive definiteness of solutions and the normalization condition of the probability density unlike the Hermite DAF-based method. In the first order of approximation over time the method proposed makes it possible to obtain solutions with higher accuracy, especially in a steady state and in the region where the solution is close to discontinuity.
The necessity of choosing a sufficiently dense uniform grid can be classified as a limitation of the proposed method. However, this can be avoided if the proposed scheme is transferred onto a non-uniform grid.

Mean field approximation was applied to the study of the evolution of the system describing the well-known model of autocatalytic chemical reaction with spatially correlated multiplicative noise. In this paper the region of parameters, in which the Turing bifurcation arises in the deterministic system, was considered.
As a result of the numerical study of NSCFPE solutions for a stochastic spatially extended brusselator we found that only unimodal probability distribution (ordered phase) can be observed at low noise intensity. The increase of noise intensity leads to the occurrence of transient bimodality (disordered phase) in the ordered phase. Further growth of noise intensity disrupts the ordering to an even greater extent: "intermittency" of unimodal and bimodal distribution, that is "phase intermittency" is observed which "swings" the ordered state. The behavior of variance over time, the most probable and mean of the function defining the system state in case of increasing the external noise intensity and the bifurcation parameter has been studied. It was shown that the most probable has the discontinuity of the first kind when transient bimodality disappears.

The behavior of the order parameter of the system under consideration was studied. It was shown that some statistical characteristics of the order parameter and the functions defining the system state behave in a similar way in the steady state. Thus, in the vicinity of the bifurcation point the greater the external noise intensity, the greater the variance in the steady state. Simultaneously the increase of noise intensity leads to the increase of difference between the mean and the appropriate most probable in the steady state. It was shown that transient bimodality occurs in the ordered phase when noise intensity values are close to the values corresponding to the transition from bimodal density of steady-state order parameter probability distribution to unimodal one.

\begin{center}
ACKNOWLEDGMENTS
\end{center}

The study has been supported by the Ministry of Education and Science of the
Russian Federation, state work specification for the years 2014-2016,
project No.608, grants of the Russian Foundation for Basic Research
13-01-970050 r\_povolzhie\_a, 13-01-97001 r\_povolzhie\_a, 14-02-97030 r\_povolzhie\_a, and grants of Russian Science Foundation ¹ 14-11-00290, 14-12-00472, 14-22-00111.

\begin{center}
\textbf{APPENDIX A}
\end{center}

\bigskip

A rectangular spatial mesh $\omega _{h} = (\{ih_{1} ,jh_{2} \})$ was chosen
for problem (\ref{eq24}). Here $i,j\,\,\,(i = 0,1,...,I;j = 0,1,...,J)$ and $h_{1}
,h_{2} $ are indexes of mesh nodes and steps respectively, and $\omega
_{\tau}  $ is a temporal mesh with a step $\tau $ on the interval $0 \le t
\le T.$ We associate the discrete function $w_{ij}^{k} $ defined on the mash
$\omega _{h} \times \omega _{\tau}  $ with the continuous function $w(x_{1}
,x_{2} ,t)$.

We choose the initial conditions

\[
w_{0} = w(x_{1} ,x_{2} ,0) = {\frac{{1}}{{2\pi \sqrt {\theta _{1} \theta
_{2}}} } }\exp {\left\{ { - {\frac{{\left( {x_{1} - x_{10}}
\right)^{2}}}{{2\theta _{1}}} } - {\frac{{\left( {x_{2} - x_{20}}
\right)^{2}}}{{2\theta _{2}}} }} \right\}},
\quad
x_{10} = A,\,\,x_{20} = B / A
\]

\noindent
and boundary conditions $w(x_{1} ,x_{2} ,t) \to 0$ if $x_{1} \to \infty
,x_{2} \to \infty $.

A locally one-dimensional scheme (\ref{eq13})-(\ref{eq15}) for problem (\ref{eq24}) has the form:

\[
{\frac{{w^{k + \alpha / n} - w^{k + (\alpha - 1) / n}}}{{\tau}} } - \Lambda
_{\alpha}  w = 0,\,\,\,w^{0} = w_{0} ,\,\,\,\alpha = 1,2;\,\,n = 2.
\]

\[
\Lambda _{\alpha}  w = (a_{\alpha}  (q_{\alpha}  w)_{\bar {x}_{\alpha}}
)_{x_{\alpha}}   ,\,\,
\]

\[
q_{1} = \exp \left( {{\frac{{a_{0}}} {{x_{1}}} } + a_{1} \ln {\left| {x_{1}
} \right|}\,\, - {\frac{{x_{1} x_{2}}} {{\theta _{1}}} }} \right),
q_{2} = \exp \left( { - b_{1} (x_{1} )x_{2} + b_{2} (x_{1} )x_{2}^{2}}
\right),
\]

\[
a_{1,i} = {\left\{ {{\frac{{1}}{{2\theta _{1}}} }{\left[ {\exp \left(
{{\frac{{a_{0}}} {{x_{1,i}}} } - {\frac{{x_{1,i} x_{2}}} {{\theta _{1}}} }}
\right)x_{1,i}^{a_{1} - 2} + \exp \left( {{\frac{{a_{0}}} {{x_{1,i - 1}}} }
- {\frac{{x_{1,i - 1} x_{2}}} {{\theta _{1}}} }} \right)x_{1,i - 1}^{a_{1} -
2}}  \right]}} \right\}}^{ - 1},x_{1,i} = ih_{1} ,
\]

\[
a_{2,j} = {\left\{ {{\frac{{1}}{{2\theta _{2} x_{1}^{2}}} }{\left[ {\exp
\left( { - b_{1} (x_{1} )x_{2,j} + b_{2} (x_{1} )x_{2,j}^{2}}  \right) +
\exp \left( { - b_{1} (x_{1} )x_{2,j - 1} + b_{2} (x_{1} )x_{2,j - 1}^{2}}
\right)} \right]}} \right\}}^{ - 1},x_{2,j} = jh_{2} .
\]

Here the notations are introduced:

\[
\begin{array}{l}
 a_{0} = {\frac{{A + D_{1} E(x_{1} \vert x_{2} ,t)}}{{\theta _{1}
}}},\,\,\,\,\,\,\,\,\,\,\,\,\,\,\,\,\,\,\,\,\,\,\,\,\,\,\,\,\,a_{1} =
{\frac{{B + 1 + \theta _{1} + D_{1}}} {{\theta _{1}}} }, \\
 b_{1} (x_{1} ) = {\frac{{B}}{{\theta _{2} x_{1}}} } + {\frac{{D_{2} E(x_{2}
\vert x_{1} ,t)}}{{\theta _{2} x_{1}^{2}
}}},\,\,\,\,\,\,\,\,\,\,\,\,\,\,\,b_{2} (x_{1} ) = {\frac{{x_{1}^{2} + D_{2}
}}{{2\theta _{2} x_{1}^{2}}} }. \\
 \end{array}
\]

On the boundary:

\[
w_{1j}^{k} = w_{0j}^{k} \sim 0,\,\,w_{i\,1}^{k} = w_{i\,0}^{k} \sim
0,	w_{Ij}^{k} = w_{I - 1j}^{k} \sim 0,\,\,w_{i\,J}^{k} = w_{i\,J - 1}^{k}
\sim 0.
\]

At the transition to a semilayer:

\[
 - B_{i} w_{i + 1j}^{k + 1 / 2} + C_{i} w_{i\,j}^{k + 1 / 2} - A_{i} w_{i -
1\,j}^{k + 1 / 2} = w_{i\,j}^{k} .
\]

\[
A_{0} = 0;\,B_{0} = 1;\,C_{0} = 1;F_{0} = 0.
\quad
A_{I} = 1;\,B_{I} = 0;\,C_{I} = 1;F_{I} = 0.
\]

\[
\begin{array}{l}
 A_{i} = {\frac{{\tau \theta _{1}}} {{h_{1}^{2}}} }{\frac{{{\left| {x_{1} -
h_{1}}  \right|}^{2}}}{{1 + \exp {\left\{ { - {\frac{{a_{0} h_{1}}} {{x_{1}
(x_{1} - h_{1} )}}} - {\frac{{x_{2} h_{1}}} {{\theta _{1}}} } + (a_{1} -
2)\ln {\left| {{\frac{{x_{1}}} {{x_{1} - h_{1}}} }} \right|}} \right\}}}}},
\\
 B_{i} = {\frac{{\tau \theta _{1}}} {{h_{1}^{2}}} }{\frac{{{\left| {x_{1} +
h_{1}}  \right|}^{2}}}{{1 + \exp {\left\{ {{\frac{{a_{0} h_{1}}} {{x_{1}
(x_{1} + h_{1} )}}} + {\frac{{x_{2} h_{1}}} {{\theta _{1}}} } + (a_{1} -
2)\ln {\left| {{\frac{{x_{1}}} {{x_{1} + h_{1}}} }} \right|}} \right\}}}}},
\\
 C_{i} = 1 + \\
 + {\frac{{\tau \theta _{1} x_{1}^{2}}} {{h_{1}^{2}}} }{\frac{{1}}{{1 + \exp
{\left\{ { - {\frac{{a_{0} h_{1}}} {{x_{1} (x_{1} + h_{1} )}}} -
{\frac{{x_{2} h_{1}}} {{\theta _{1}}} } + (a_{1} - 2)\ln {\left|
{{\frac{{x_{1} + h_{1}}} {{x_{1}}} }} \right|}} \right\}}}}} + \\
 {\frac{{\tau \theta _{1} x_{1}^{2}}} {{h_{1}^{2}}} }{\frac{{1}}{{1 + \exp
{\left\{ {{\frac{{a_{0} h_{1}}} {{x_{1} (x_{1} - h_{1} )}}} + {\frac{{x_{2}
h_{1}}} {{\theta _{1}}} } + (a_{1} - 2)\ln {\left| {{\frac{{x_{1} - h_{1}
}}{{x_{1}}} }} \right|}} \right\}}}}} \\
 \end{array}
\]

At the transition to a whole layer:

\[
 - B_{j} w_{i\,j + 1}^{k + 1} + C_{j} w_{i\,j}^{k + 1} - A_{j} w_{i\,j -
1}^{k + 1} = w_{i\,j}^{k + 1 / 2} .
\]

\[
A_{0} = 0;\,B_{0} = 1;\,C_{0} = 1;F_{0} = 0.
\quad
A_{J} = 1;\,B_{J} = 0;\,C_{J} = 1;F_{J} = 0.
\]

\[
\begin{array}{l}
 A_{j} = {\frac{{\tau \theta _{2} x_{1}^{2}}} {{h_{2}^{2}}} }{\frac{{1}}{{1
+ \exp {\left\{ { - b_{1} h_{2} + 2b_{2} h_{2} (x_{2} - 0.5h_{2} )}
\right\}}}}}, \\
 B_{j} = {\frac{{\tau \theta _{2} x_{1}^{2}}} {{h_{2}^{2}}} }{\frac{{1}}{{1
+ \exp {\left\{ {b_{1} h_{2} - 2b_{2} h_{2} (x_{2} + 0.5h_{2} )}
\right\}}}}}, \\
 C_{j} = 1 + {\frac{{\tau \theta _{2} x_{1}^{2}}} {{h_{1}^{2}}} }{\left[
{{\frac{{1}}{{1 + \exp {\left\{ { - b_{1} h_{2} + 2b_{2} h_{2} (x_{2} +
0.5h_{2} )} \right\}}}}} + {\frac{{1}}{{1 + \exp {\left\{ {b_{1} h_{2} -
2b_{2} h_{2} (x_{2} - 0.5h_{2} )} \right\}}}}}} \right]}. \\
 \end{array}
\]

\bigskip

\begin{center}
\textbf{APPENDIX B}
\end{center}

\bigskip

The functions introduced in Eq.(\ref{eq25}):

\[
 \lambda _{\mu}  ({\mathbf{k}}) = {\frac{{\alpha ({\mathbf{k}})}}{{2}}}\pm
\sqrt {{\frac{{\alpha ^{2}({\mathbf{k}})}}{{4}}} - \beta ({\mathbf{k}})} ,
\]
\[
\alpha ({\mathbf{k}}) = B - 1 - A^{2} - (D_{1} + D_{2} )k^{2},\\
\beta({\mathbf{k}}) = [A^{2}D_{1} - (B - 1)D_{2} ]k^{2} + D_{1} D_{2} k^{4} + A^{2}, \\
\]

\[
\Omega _{1} ({\mathbf{k}}_{u} ,{\mathbf{k}'}_{u} ,{\mathbf{k}}_{s},{\mathbf{z}}(\tau )) =
{\sum\limits_{\mu ,\varphi}  {\eta _{\varphi
}^{(\mu )} ({\mathbf{k}}_{u} ,{\mathbf{k}'}_{u} )z_{\varphi ,{\mathbf{k}}u - {\mathbf{k}'}u}}}   -
{\sum\limits_{{\mathbf{k}'}s}
{{\sum\limits_{\mu ,\varphi ,{\varphi} '} {A_{\varphi {\varphi} '}^{(\mu )}
({\mathbf{k}'}_{u} ,{\mathbf{k}}_{s} ,{\mathbf{k}}_{u} )z_{\varphi
,{\mathbf{k}}u - {\mathbf{k}}s} z_{{\varphi} ',{\mathbf{k}}s - {\mathbf{k}'}u}}} } } ,
\]

\[
\Omega _{11} ({\mathbf{k}}_{u} ,{\mathbf{k}'}_{u} ,{\mathbf{k}''}_{u}
,{\mathbf{k}}_{s} ,{\mathbf{z}}(\tau )) = \delta ({\mathbf{k}}_{u} ,{\mathbf{k}'}_{u} ,{\mathbf{k}''}_{u} ) - {\sum\limits_{\mu ,\varphi}  {( -
1)^{\varphi} \nu _{\varphi} ^{(\mu )} ({\mathbf{k}}_{u} ,{\mathbf{k}'}_{u} ,{\mathbf{k}''}_{u} )z_{\varphi ,{\mathbf{k}}u - {\mathbf{k}'}u - {\mathbf{k}''}u}}} ,
\]

\[
\eta _{\varphi} ^{(\mu )} ({\mathbf{k}}_{u} ,{\mathbf{k}'}_{u} ) = ( -
1)^{\varphi} [O_{1}^{(1)} ({\mathbf{k}'}_{u} )O_{\varphi} ^{ * (1)} ({\mathbf{k}}_{u} ) - \beta _{\varphi} ^{(\mu )}
 ({\mathbf{k}}_{u} ,{\mathbf{k}'}_{u} )],
\]

\[
A_{\varphi {\varphi} '}^{(\mu )} ({\mathbf{k}'}_{u} ,{\mathbf{k}}_{s}
,{\mathbf{k}}_{u} ) = ( - 1)^{\varphi + {\varphi} '}O_{1}^{(1)} ({\mathbf{k}'}_{u} )\varepsilon _{\varphi {\varphi} '}^{(\mu )} ({\mathbf{k}}_{s},{\mathbf{k}'}_{u} ),
\]

\[
\zeta _{\varphi {\varphi} '}^{(\mu )} ({\mathbf{k}}_{s} ,{\mathbf{k}}_{u}
) = ( - 1)^{\varphi + {\varphi} '}A\varepsilon _{\varphi {\varphi} '}^{(\mu
)} ({\mathbf{k}}_{s} ,{\mathbf{k}'}_{u} ),
\]

\[
\varepsilon _{\varphi {\varphi} '}^{(\mu )} ({\mathbf{k}}_{s} ,{\mathbf{k}'}_{u} ) = {\frac{{O_{1}^{(\mu )} ({\mathbf{k}}_{s} )}}{{\lambda _{\mu
} ({\mathbf{k}}_{s} )}}}O_{\varphi} ^{ * (\mu )} ({\mathbf{k}}_{s}
)O_{{\varphi} '}^{ * (1)} ({\mathbf{k}}_{u} ),
\]

\[
\beta _{\varphi} ^{(\mu )} ({\mathbf{k}}_{u} ,{\mathbf{k}'}_{u} ) =
{\frac{{O_{1}^{ * (1)} ({\mathbf{k}}_{u} ) - O_{2}^{ * (1)} ({\mathbf{k}}_{u} )}}{{\lambda _{\mu}  ({\left| {{\mathbf{k}}_{u} - {\mathbf{k}'}_{u}}  \right|})}}}O_{\varphi} ^{ * (\mu )} ({\left| {{\mathbf{k}}_{u} - {\mathbf{k}'}_{u}}  \right|})\sigma ^{1\mu} ({\mathbf{k}}_{u}
,{\left| {{\mathbf{k}}_{u} - {\mathbf{k}'}_{u}}  \right|}),
\]

\[
 \nu _{\varphi} ^{(\mu )} ({\mathbf{k}}_{u} ,{\mathbf{k}'}_{u} ,{\mathbf{k}''}_{u} ) =
 \beta _{\varphi} ^{(\mu )} ({\mathbf{k}}_{u} ,{\mathbf{k}'}_{u} )O_{1}^{(1)} ({\mathbf{k}''}_{u} ) +
 \]
\[
 O_{\varphi} ^{ * (1)} ({\mathbf{k}}_{u} ){\frac{{O_{1}^{(\mu )} ({\left|
{{\mathbf{k}'}_{u} + {\mathbf{k}''}} \right|}_{u} )}}{{\lambda _{\mu}
({\left| {{\mathbf{k}'}_{u} + {\mathbf{k}''}_{u}}
\right|})}}}[O_{1}^{ * (\mu )} ({\left| {{\mathbf{k}'}_{u} + {\mathbf{k}''}_{u}}  \right|}) -
 \]
\[
O_{2}^{ * (\mu )} ({\left| {{\mathbf{k}'}_{u} +
{\mathbf{k}''}_{u}}  \right|})][2AO_{1}^{(1)} ({\mathbf{k}'}_{u}
)O_{2}^{(1)} ({\mathbf{k}''}_{u} ) + {\frac{{B}}{{A}}}O_{1}^{(1)} ({\mathbf{k}'}_{u} )O_{1}^{(1)} ({\mathbf{k}''}_{u} )], \\
\]

\[
\delta ({\mathbf{k}}_{u} ,{\mathbf{k}'}_{u} ,{\mathbf{k}''}_{u} ) =
[O_{1}^{ * (1)} ({\mathbf{k}}_{u} ) - O_{2}^{ * (1)} ({\mathbf{k}}_{u}
)][2AO_{1}^{(1)} ({\mathbf{k}'}_{u} )O_{2}^{(1)} ({\mathbf{k}''}_{u} )
+ {\frac{{B}}{{A}}}O_{1}^{(1)} ({\mathbf{k}'}_{u} )O_{1}^{(1)} ({\mathbf{k}''}_{u} )]\delta ({\mathbf{k}}_{u} -
{\mathbf{k}'}_{u} - {\mathbf{k}''}_{u} ),
\]

\[
 \omega ({\mathbf{k}}_{u} ,{\mathbf{k}'}_{u} ,{\mathbf{k}''}_{u}
,{\mathbf{k}'''}_{u} ) = [O_{1}^{ * (1)} ({\mathbf{k}}_{u} ) - O_{2}^{ *
(1)} ({\mathbf{k}}_{u} )]O_{1}^{(1)} ({\mathbf{k}'}_{u} )O_{1}^{(1)}
({\mathbf{k}''}_{u} )O_{2}^{(1)} ({\mathbf{k}'''}_{u} )\delta ({\mathbf{k}}_{u} - {\mathbf{k}'}_{u} - {\mathbf{k}''}_{u} -
{\mathbf{k}'''}_{u} ) -
\]
\[
 {\sum\limits_{{\mathbf{k}}s,\mu}  {{\left\{ {{\frac{{[O_{1}^{ * (1)} ({\mathbf{k}}_{u} ) - O_{2}^{ * (1)} ({\mathbf{k}}_{u} )]}}{{\lambda _{\mu}
({\mathbf{k}}_{s} )}}}\sigma ^{1\mu} ({\mathbf{k}'}_{u} ,{\mathbf{k}}_{s} )[O_{1}^{ * (\mu )} ({\mathbf{k}}_{s} ) - O_{2}^{ * (\mu )} ({\mathbf{k}}_{s} )]} \right.}}} \times
\]
\[
 {\left. {[2AO_{1}^{(1)} ({\mathbf{k}'''}_{u} )O_{2}^{(1)} ({\mathbf{k}''}_{u} ) + {\frac{{B}}{{A}}}O_{1}^{(1)}
  ({\mathbf{k}'''}_{u}
)O_{1}^{(1)} ({\mathbf{k}''}_{u} )]\delta ({\mathbf{k}}_{u} - {\mathbf{k}'}_{u} - {\mathbf{k}}_{s} )\delta ({\mathbf{k}}_{s} - {\mathbf{k}''}_{u} - {\mathbf{k}'''}_{u} )} \right\}} ,
\]

\[
\sigma ^{1\mu} ({\mathbf{k}'}_{u} ,{\mathbf{k}}_{s} ) = 2AO_{1}^{(1)}
({\mathbf{k}'}_{u} )O_{2}^{(\mu )} ({\mathbf{k}}_{s} ) +
2{\frac{{B}}{{A}}}O_{1}^{(1)} ({\mathbf{k}'}_{u} )O_{1}^{(\mu )} ({\mathbf{k}}_{s} ) + 2AO_{1}^{(\mu )} ({\mathbf{k}}_{s} )O_{2}^{(1)} ({\mathbf{k}'}_{u} ),
\]

\[
\begin{array}{l}
 {\mathbf{O}}^{(\mu )} ({\mathbf{k}}) = \left( {{\begin{array}{*{20}c}
 {( - A^{2} - D_{2} k^{2} - \lambda _{\mu}  ({\mathbf{k}})) / B} \hfill \\
 {1} \hfill \\
\end{array}}}  \right), \\
 {\mathbf{O}}^{ * (\mu )} ({\mathbf{k}}) = \left(
{{\begin{array}{*{20}c}
 {( - 1)^{\mu} O_{2}^{({\mu} ')} ({\mathbf{k}}) / [O_{2}^{(1)} ({\mathbf{k}})O_{1}^{(2)} ({\mathbf{k}}) - O_{2}^{(2)}
  ({\mathbf{k}})O_{1}^{(1)}
({\mathbf{k}})]} \hfill \\
 {( - 1)^{{\mu} '}O_{1}^{({\mu} ')} ({\mathbf{k}}) / [O_{2}^{(1)} ({\mathbf{k}})O_{1}^{(2)} ({\mathbf{k}}) - O_{2}^{(1)}
  ({\mathbf{k}})O_{1}^{(1)} ({\mathbf{k}})]} \hfill \\
\end{array}}}  \right), {\rm {if}} \,\mu = 1,{\mu} ' = 2;{\rm {if}} \,\mu = 2,{\mu} ' = 1. \\
 \end{array}
\]

\bigskip

\begin{center}
\textbf{APPENDIX C}
\end{center}

\bigskip

The correlators from Eq. (\ref{eq26}):

\[
K[{\frac{{\partial F_{{\mathbf{k}}u} (\tau )}}{{\partial \xi _{{\mathbf{q}}u}^{(1)}}} },F_{{\mathbf{q}}u} ({t}')] = {\sum\limits_{\varphi}  {\eta
_{\varphi}  ({\mathbf{k}}_{u} ,{\mathbf{q}}_{u} )O_{\varphi} ^{ * (1)}
({\mathbf{q}}_{u} )p_{\varphi} ^{(0)} g_{\varphi \varphi}  (\vert {\mathbf{k}}_{u} - {\mathbf{q}}_{u} \vert )}}
 \delta _{{\mathbf{k}}u - {\mathbf{q}}u,{q}u} \delta (\tau - {t}') +
\]
\[
{\sum\limits_{\varphi}  {[\nu _{\varphi}  ({\mathbf{k}}_{u} ,{\mathbf{q}}_{u} ,{\mathbf{k}}_{u} - 2{\mathbf{q}}_{u} ) + \nu _{\varphi}  ({\mathbf{k}}_{u} ,{\mathbf{k}}_{u} - 2{\mathbf{q}}_{u} ,{\mathbf{q}}_{u}
)]O_{\varphi} ^{ * (1)} ({\mathbf{q}}_{u} )p_{\varphi} ^{(0)} g_{\varphi
\varphi}  (\vert {\mathbf{q}}_{u} \vert )\xi _{{\mathbf{k}}u - 2{\mathbf{q}}u} \delta (\tau - {t}') +}}
\]
\[
{\sum\limits_{\varphi}  {\eta _{\varphi}  ({\mathbf{k}}_{u} ,{\mathbf{q}}_{u} )\eta _{\varphi}
 ({\mathbf{q}}_{u},2{\mathbf{q}}_{u} - {\mathbf{k}}_{u} )g_{\varphi \varphi}  (\vert {\mathbf{k}}_{u} - {\mathbf{q}}_{u}
\vert )\xi _{2{\rm {\bf q}}u - {\rm {\bf k}}u}}}  \delta (\tau - {t}') +
\]
\[
{\sum\limits_{\varphi ,{\mathbf{q}'}u} {\eta _{\varphi}
 ({\mathbf{k}}_{u} ,{\mathbf{q}}_{u} )\nu _{\varphi}  ({\mathbf{q}}_{u} ,{\mathbf{q}'}_{u} ,2{\mathbf{q}}_{u} -
  {\mathbf{k}}_{u} - {\mathbf{q}'}_{u}
)g_{\varphi \varphi}  (\vert {\mathbf{k}}_{u} - {\mathbf{q}}_{u} \vert
)\xi _{{\mathbf{q}'}u} \xi _{2{\mathbf{q}}u - {\mathbf{k}}u - {\mathbf{q}'}u}}}  \delta (\tau - {t}') +
\]
\[
{\sum\limits_{\varphi ,{\mathbf{q}'}u,{\mathbf{q}''}u} {[\nu
_{\varphi}  ({\mathbf{k}}_{u} ,{\mathbf{q}}_{u} ,{\mathbf{k}}_{u} -
 2{\mathbf{q}}_{u} + {\mathbf{q}'}_{u} + {\mathbf{q}''}_{u} ) + \nu
_{\varphi}  ({\mathbf{k}}_{u} ,{\mathbf{k}}_{u} - 2{\mathbf{q}}_{u} +
{\mathbf{q}'}_{u} + {\mathbf{q}''}_{u} ,{\mathbf{q}}_{u} )]\times}}
\]
\[
\nu _{\varphi}  ({\mathbf{q}}_{u}
,{\mathbf{q}'}_{u} ,{\mathbf{q}''}_{u} )g_{\varphi \varphi}  (\vert
{\mathbf{q}}_{u} - {\mathbf{q}'}_{u} - {\mathbf{q}''}_{u} \vert )\xi
_{{\mathbf{q}'}u} \xi _{{\mathbf{q}''}u} \xi _{{\mathbf{k}}u -
 2{\mathbf{q}}u + {\mathbf{q}'}u + {\mathbf{q}''}u} \delta (\tau - {t}') +
\]
\[
{\sum\limits_{\varphi ,{\mathbf{q}'}u} {\eta _{\varphi}  ({\mathbf{q}}_{u} ,{\mathbf{q}'}_{u} )[\nu _{\varphi}
 ({\mathbf{k}}_{u} ,{\mathbf{k}}_{u} - 2{\mathbf{q}}_{u} + {\mathbf{q}'}_{u} ,{\mathbf{q}}_{u})
 + \nu _{\varphi}  ({\mathbf{k}}_{u} ,{\mathbf{q}}_{u} ,{\mathbf{k}}_{u}
- 2{\mathbf{q}}_{u} + {\mathbf{q}'}_{u} )]}} \times
\]
\[
g_{\varphi \varphi}
(\vert {\mathbf{q}}_{u} - {\mathbf{q}'}_{u} \vert )
\xi _{{\mathbf{q}'}u} \xi _{{\mathbf{k}}u - 2{\mathbf{q}}u + {\mathbf{q}'}u} \delta(\tau - {t}'),
\]

\[
K[F_{{\mathbf{k}}u} (\tau ),F_{{\mathbf{q}}u} ({t}')] =
{\sum\limits_{\varphi}  {\left( {O_{\varphi} ^{ * (1)} ({\mathbf{k}}_{u} )}
\right)^{2}\left( {p_{\varphi} ^{(0)}}  \right)^{2}g_{\varphi \varphi}
(\vert {\mathbf{k}}_{u} \vert )}} \delta _{{\mathbf{k}}u,{\mathbf{q}}u}
\delta (\tau - {t}') +
\]
\[
{\sum\limits_{\varphi}  {\eta _{\varphi}  ({\mathbf{k}}_{u} ,{\mathbf{k}}_{u} - {\mathbf{q}}_{u} )
O_{\varphi} ^{ * (1)} ({\mathbf{q}}_{u}) p_{\varphi} ^{(0)} g_{\varphi \varphi}  (\vert {\mathbf{q}}_{u} \vert )\xi
_{{\mathbf{k}}u - {\mathbf{q}}u}}}  \delta (\tau - {t}') +
\]
\[
{\sum\limits_{\varphi}  {\eta _{\varphi}  ({\mathbf{q}}_{u} ,
{\mathbf{q}}_{u} - {\mathbf{k}}_{u} )O_{\varphi} ^{ * (1)} ({\mathbf{k}}_{u}
)p_{\varphi} ^{(0)} g_{\varphi \varphi}  (\vert {\mathbf{k}}_{u} \vert )\xi
_{{\mathbf{q}}u - {\mathbf{k}}u}}}  \delta (\tau - {t}') +
\]
\[
{\sum\limits_{\varphi ,{\mathbf{q}'}u} {\eta _{\varphi}
  ({\mathbf{k}}_{u} ,{\mathbf{k}}_{u} - {\mathbf{q}}_{u} + {\mathbf{q}'}_{u} )\eta
_{\varphi}  ({\mathbf{q}}_{u} ,{\mathbf{q}'}_{u} )g_{\varphi \varphi}
(\vert {\mathbf{q}}_{u} - {\mathbf{q}'}_{u} \vert )
\xi _{{\mathbf{q}'}u}}}  \xi _{{\mathbf{k}}u - {\mathbf{q}}u + {\mathbf{q}'}u}
\delta (\tau - {t}') +
\]
\[
{\sum\limits_{\varphi ,{\mathbf{k}'}u,{\mathbf{q}'}u} {\nu
_{\varphi}  ({\mathbf{k}}_{u} ,{\mathbf{k}'}_{u} ,{\mathbf{k}}_{u} -
{\mathbf{k}'}_{u} - {\mathbf{q}}_{u} + {\mathbf{q}'}_{u} )\eta
_{\varphi}  ({\mathbf{q}}_{u} ,{\mathbf{q}'}_{u} )g_{\varphi \varphi}
(\vert {\mathbf{q}}_{u} - {\mathbf{q}'}_{u} \vert )
\xi _{{\mathbf{k}'}u} \xi _{{\mathbf{q}'}u} \xi _{{\mathbf{k}}u - {\mathbf{k}'}u -
{\mathbf{q}}u + {\mathbf{q}'}u} \delta (\tau - {t}') +}}
\]
\[
{\sum\limits_{\varphi ,{\mathbf{q}'}u} {\nu _{\varphi}
({\mathbf{q}}_{u} ,{\mathbf{q}'}_{u} ,{\mathbf{q}}_{u} - {\mathbf{q}'}_{u} -
{\mathbf{k}}_{u} )O_{\varphi} ^{ * (1)} ({\mathbf{k}}_{u} )p_{\varphi
}^{(0)} g_{\varphi \varphi}  (\vert {\mathbf{k}}_{u} \vert )
\xi _{{\mathbf{q}'}u}}}  \xi _{{\mathbf{q}}u - {\mathbf{q}'}u - {\mathbf{k}}u}
\delta (\tau - {t}') +
\]
\[
{\sum\limits_{\varphi ,{\mathbf{k}'}u,{\mathbf{q}'}u} {\nu
_{\varphi}  ({\mathbf{q}}_{u} ,{\mathbf{q}'}_{u} ,{\mathbf{q}}_{u} -
{\mathbf{q}'}_{u} - {\mathbf{k}}_{u} + {\mathbf{k}'}_{u} )\eta
_{\varphi}  ({\mathbf{k}}_{u} ,{\mathbf{k}'}_{u} )}} g_{\varphi \varphi
} (\vert {\mathbf{k}}_{u} - {\mathbf{k}'}_{u} \vert )
\xi _{{\mathbf{k}'}u} \xi _{{\mathbf{q}'}u} \xi _{{\mathbf{q}}u - {\mathbf{q}'}u -
{\mathbf{k}}u + {\mathbf{k}'}u} \delta (\tau - {t}') +
\]
\[
{\sum\limits_{\varphi ,{\mathbf{k}'}u,{\mathbf{q}'}u,{\mathbf{q}''}u} {\nu _{\varphi}
 ({\mathbf{k}}_{u} ,{\mathbf{k}'}_{u} ,{\mathbf{k}}_{u} - {\mathbf{k}'}_{u} - {\mathbf{q}}_{u} +
 {\mathbf{q}'}_{u} + {\mathbf{q}''}_{u} )\nu _{\varphi}  ({\mathbf{q}}_{u} ,
 {\mathbf{q}'}_{u} ,{\mathbf{q}''}_{u} )\times}}
\]
\[
 g_{\varphi \varphi}  (\vert {\mathbf{q}}_{u} - {\mathbf{q}'}_{u}
- {\mathbf{q}''}_{u} \vert )\xi _{{\mathbf{k}'}u}
\xi _{{\mathbf{q}'}u} \xi _{{\mathbf{q}''}u} \xi _{{\mathbf{k}}u - {\mathbf{k}'}u
- {\mathbf{q}}u + {\mathbf{q}'}u + {\mathbf{q}''}u} \delta (\tau -
{t}').
\]


\begin{thebibliography}{99}
\bibitem{Lind1}B. Lindnera, J. Garc\'{\i}a-Ojalvo, A. Neimand, and L. Schimansky-Geier, Phys. Rep. \textbf{392}, 321 (2004).
\bibitem{Iba2}M. Iba\~{n}es, J. Garc\'{\i}a-Ojalvo, R. Toral, and J. M. Sancho, Phys. Rev. E \textbf{60}, 3597 (1999).
\bibitem{Buc3}J. Buceta, M. Iba\~{n}es, J. M. Sancho, and K. Lindenberg, Phys. Rev. E \textbf{67}, 021113 (2003).
\bibitem{Carr4}O. Carrillo, M. Iba\~{n}es, J. Garc\'{\i}a-Ojalvo, J. Casademunt, and J. M. Sancho, Phys. Rev. E \textbf{67}, 046110 (2003).
\bibitem{Zaik5}A. A. Zaikin, J. Garc\'{\i}a-Ojalvo, and L. Schimansky-Geier, Phys. Rev. E \textbf{60}, R6275 (1999).
\bibitem{Mull6}R. M\"{u}ller, K. Lippert, A. K\"{u}hnel, and U. Behn, Phys. Rev. E \textbf{56}, 2658 (1997).
\bibitem{Carr7}O. Carrillo, M. Iba\~{n}es, and J.M. Sancho, Fluct. Noise Lett. \textbf{2}, L1 (2002).
\bibitem{Land8}P. S. Landa, A.A. Zaikin, L. Schimansky-Geier, Chaos, Solitons and Fractals \textbf{9}, 1367 (1998).
\bibitem{Bro9}C. Van den Broeck, J. M. R. Parrondo, R. Toral, and R. Kawai, Phys. Rev. E \textbf{55}, 4084 (1997).
\bibitem{Buce10}J. Buceta, J. M. R. Parrondo, and F. Javier de la Rubia, Phys. Rev. E \textbf{63}, 031103 (2001).
\bibitem{Zhan11}D. S. Zhang, G. W. Wei, and D. J. Kouri, Phys. Rev. E \textbf{56}, 1197 (1997).
\bibitem{Zhang12}D. S. Zhang, G. W. Wei, D. J. Kouri, and D. K. Hoffman, J. Chem. Phys. \textbf{106}, 5216 (1997).
\bibitem{Drozd13}A. N. Drozdov and M. Morillo, Phys. Rev. E \textbf{54}, 931 (1996).
\bibitem{Chen14}H. Chen, J. Duan, and Ch. Zhang, Acta Mathematica Scientia \textbf{32}, 1391 (2012).
\bibitem{Kum15}P. Kumar and S. Narayanan, S$\bar{a}$dhan$\bar{a}$ \textbf{31}, 445 (2006).
\bibitem{Camp16}F. Campillo, M. Joannides, and I. Larramendy-Valverde, Mathematics and Computers in Simulation \textbf{99}, 37 (2014).
\bibitem{Wei17}G. W. Wei, J. Chem. Phys. \textbf{110}, 8930 (1999); J. Phys. A: Math. Gen. \textbf{33}, 4935 (2000).
\bibitem{Ott18}D. L. Otten and P. Vedula, J. Stat. Mech. P09031 (2011).
\bibitem{Fox19}R. O. Fox and P. Vedula, Ind. Eng. Chem. Res. \textbf{49}, 5174 (2010).
\bibitem{Kawam20}Y. Kawamura, Phys. Rev. E \textbf{76}, 047201 (2007).
\bibitem{Wehn21}M. H. Wehner and W. G. Wolfer, Phys. Rev. A \textbf{35}, 1795 (1987).
\bibitem{Hak22}H. Haken, Z. Phys. B \textbf{24}, 321 (1976).
\bibitem{Kamp23}N. G. van Kampen, J. Stat. Phys. \textbf{17}, 71 (1977).
\bibitem{Tomit24}H. Tomita, A. Ito, and H. Kidachi, Prog. Theor. Phys. \textbf{56}, 786 (1976).
\bibitem{Mor25}D. Moroni, B. Rotenberg, J.-P. Hansen, S. Succi, and S. Melchionna, Phys. Rev. E \textbf{73}, 066707 (2006).
\bibitem{Erm26} D. L. Ermak and H. Buckholtz, J. Comput. Phys. \textbf{35}, 169 (1980).
\bibitem{Kur27}S. E. Kurushina, V. V. Maximov, and Yu.M. Romanovskii, Phys. Rev. E \textbf{86}, 011124 (2012).
\bibitem{Prig28}I. Prigogine and R. Lefever, J. Chem. Phys. \textbf{48}, 1695 (1968).
\bibitem{Shim29}H. Shimizu and T. Yamada, Prog.Theor. Phys. \textbf{47}, 350 (1972).
\bibitem{Shim30}H. Shimizu, Prog. Theor. Phys. \textbf{52}, 329 (1974).
\bibitem{Desai31}R. C. Desai and R. Zwanzig, J. Stat. Phys. \textbf{19}, 1 (1978).
\bibitem{Horst32}W. Horsthemke and M. Lefever, \textit{Noise-Induced Transition} (Springer, Berlin, 1984).
\bibitem{Garc33}J. Garc\'{\i}a-Ojalvo, A. M. Lacasta, J. M. Sancho, and R. Toral, Europhys. Lett. \textbf{42}, 125 (1998).
\bibitem{Strat34}R. L. Stratonovich, \textit{Topics in the Theory of Random Noise}, (Gordon and Breach, New York, London, 1963), Vol.1; (Gordon and Breach, New York, London, 1967), Vol. 2.
\bibitem{Samar35}A. A. Samarskii, USSR Computational Mathematics and Mathematical Physics \textbf{2}, 23 (1963).
\bibitem{Karet36} N. V. Karetkina, USSR Computational Mathematics and Mathematical Physics \textbf{20}, 257 (1980).
\bibitem{Samar37}A. A. Samarskii, USSR Computational Mathematics and Mathematical Physics \textbf{2}, 894 (1963).
\bibitem{Komet38}K. Kometani and H. Shimizu, J. Stat. Phys. \textbf{13}, 473 (1975).
\bibitem{Samar39}A. A. Samarskii, USSR Computational Mathematics and Mathematical Physics \textbf{3}, 572 (1963).
\bibitem{Samar40}A. A. Samarskii, USSR Computational Mathematics and Mathematical Physics \textbf{3}, 351 (1963).
\bibitem{Haken41}H. Haken, \textit{Synergetics} (Springer, Berlin, 2004).


\end{thebibliography}
\end{document}